\documentclass[11pt,onecolumn,superscriptaddress,nofootinbib,floatfix]{revtex4-1}

\pdfoutput=1
\usepackage{amsmath}
\usepackage{amsfonts}
\usepackage{amssymb}
\usepackage{mathrsfs}
\usepackage{graphicx}
\usepackage{color}
\usepackage{bm}
\usepackage{mathtools}

\DeclarePairedDelimiter\ket{\lvert}{\rangle}
\DeclarePairedDelimiterX\brake t[2]{\langle}{\rangle}{#1 \delimsize\vert #2}

\usepackage{blindtext}
\usepackage{wasysym}
\usepackage{subcaption}
\usepackage{hyperref}
\usepackage{float}
\hypersetup{colorlinks=true,allcolors=blue}
\usepackage[normalem]{ulem}
\usepackage{lineno}

\newcommand{\ee}{$\eta_{ee}~$}

\renewcommand{\tt}{$\eta_{\tau 
\tau}~$}

\newcommand{\eq}{Eq.~}
\newcommand{\eqs}{Eqs.~}

\newcommand{\fig}{Fig.~}
\newcommand{\Fig}{Fig.~}

\newcommand{\sdm}{$\Delta m^2_{21}~$}
\newcommand{\ldm}{$\Delta m^2_{31}~$}

\newcommand{\bi}{\begin{itemize}}
\newcommand{\ei}{\end{itemize}}
\newcommand{\mass}{$\mathcal{M}$~}

\begin{document}
 
\title{Neutrino oscillation measurements with KamLAND and JUNO in the presence of scalar NSI}

\author{Aman Gupta}
\email{aman.gupta@saha.ac.in(ORCID:0000-0002-7247-2424)} 
\affiliation{Theory Division, Saha Institute of
  Nuclear Physics,
  1/AF,
  Bidhannagar,
  Kolkata 700064, India} 

\affiliation{Homi  Bhabha  National  Institute,  Anushakti  Nagar,  Mumbai  400094,  India}  
\author{Debasish Majumdar\footnote{\footnotesize Present address: Department of Physics, Ramakrishna Mission Vivekananda Educational and Research Institute, Belur Math, Howrah, West Bengal 711202, India}}
\email{debasish.majumdar@gm.rkmvu.ac.in} 
\affiliation{Theory Division, Saha Institute of
  Nuclear Physics,
  1/AF,
  Bidhannagar,
  Kolkata 700064, India} 
\affiliation{Homi  Bhabha  National  Institute,  Anushakti  Nagar,  Mumbai  400094,  India}

\author{Suprabh Prakash}
\email{suprabh.prakash@vit.ac.in(ORCID:0000-0002-1529-4588)}

\affiliation{The Institute of Mathematical Sciences, C.I.T. Campus, Taramani, Chennai 600113, India} 
\affiliation{Department of Physics, School of Advanced Sciences, Vellore Institute of Technology - Chennai Campus, Chennai 600127, India}  
  

\begin{abstract}
Determination of neutrino mass ordering and precision measurement of neutrino oscillation parameters are the foremost goals of the JUNO experiment. Here, we explore the effects of scalar non-standard interactions (sNSI) on the electron anti-neutrino survival probability measured by JUNO. sNSI appear as corrections to the neutrino mass term in the Hamiltonian. We have considered the simplest scenario where there is only one NSI ($\eta_{ee}$) present in the theory. Our results show that sNSI can have a significant effect on neutrino oscillation probabilities at the medium- and long-baseline reactor experiments. We fit KamLAND data assuming non-zero sNSI in theory and find that {\it estimates of \sdm and $\theta_{12}$ from KamLAND deviate significantly from their standard best-fit values} if one assumes sNSI in the theory. $\eta_{ee} \in [-1.0, 1.0]$ is allowed by KamLAND. JUNO cannot constrain sNSI but it can robustly measure \sdm and $\theta_{12}$ even when they differ widely from their current best-fit values. {\it Our work highlights the necessity of global analysis of constraints on sNSI and standard two-flavour oscillation parameters before arduous three-flavour questions such as neutrino mass ordering or CP violation in their presence are attempted.} 

%

\end{abstract}


\maketitle

\section{Introduction}\label{sec:introduction}

The discovery of neutrino flavour oscillations \cite{Super-Kamiokande:1998kpq,Ahmad_2002,Eguchi_2003}
implies that neutrinos are massive which is one of the few observational indications  in particle physics with regard to physics beyond the Standard Model (BSM). From the evolution of the Universe to matter-antimatter asymmetry, neutrinos play very important role in cosmology and several aspects of elementary particle physics. At present, neutrino oscillation physics has entered the precision era.  In the standard three-flavour paradigm, neutrino oscillations are governed by two independent neutrino mass-squared differences, $\Delta m^2_{21}$ and $\Delta m^2_{31}$, three leptonic mixing angles: $\theta_{12}, \theta_{13}$ and $\theta_{23}$ and probably a CP-violating phase $\delta_{CP}$. Several experiments using solar, reactor, atmospheric, and accelerator neutrinos have measured these neutrino oscillation parameters with great precision \cite{ptep, Esteban:2020cvm, de_Salas_2021, Capozzi_2020}. However, there remain three principal unknowns in oscillation physics namely the octant of $\theta_{23}$, the existence and the measure of Dirac CP phase, $\delta_{\rm CP}$ and the neutrino mass hierarchy or the ordering of neutrino masses ($m_1,m_2, m_3$). In addition to this, the determination of absolute masses of neutrinos and whether neutrinos are of Dirac or Majorana type are the two fundamental problems in neutrino physics, which are yet to be answered. \\
\hspace*{1cm} The  analyses of data from various solar neutrino experiments \cite{Cleveland_1998,Kaether_2010,Bellini_2011,Abe_2011,Aharmim_2013,Ahmad_2002} suggest that $\Delta m^2_{21} (= m^2_2 - m^2_1) >0$, which means $m_2 > m_1$ where $m_1$and $m_2$ are the two of the three eigenvalues of neutrino mass matrix (the third being $m_3$). However, the ordering between $m_1$ and $m_3$ is yet to be determined. If $m_3 > m_1$ then the sign of $\Delta m^2_{31} (= m^2_3 - m^2_1$) is +ve and this scenario when $\Delta m^2_{31}$ is +ve is referred to as the normal ordering (NO) for neutrino masses  whereas, if $m_3 < m_1$ and hence $\Delta m^2_{31}$ is -ve the neutrino masses are said to follow the inverted mass ordering (IO). It is worth mentioning here that the combined analyses of all the available data from accelerator-based experiments (such as T2K \cite{Abe_2020,t2k_2021}, NO$\nu$A \cite{Acero_2019,nova_2022}) as well as those from atmospheric neutrino experiments (such as Super-Kamiokande  \cite{SK_2018} and IceCube \cite{Ice_2015}), suggest a slight preference for normal mass ordering over the inverted one \cite{Esteban:2020cvm}. Solving the problem of neutrino mass hierarchy to a significant confidence level is one of the primary goals of future accelerator-based experiments such as DUNE \cite{DUNE_1}, T2HK \cite{t2hk_2015}, and T2HKK \cite{t2hkk_2018}, but the results from these experiments will be available only by 2030 or later. Some other experiments such as atmospheric-based KM3NeT/ORCA \cite{km3net_2016} and PINGU \cite{pingu_2017} and reactor-based experiments such as JUNO \cite{JUNOAn_2016} will also address the neutrino mass ordering (NMO) problem and thus would be complementary to the accelerator-based experiments with similar aim. Currently, JUNO detector is under construction and is expected to start collecting data in 2024. This makes JUNO experiment at the frontline of determining NMO. In addition to this, JUNO is also expected to provide very accurate measurement of the solar oscillation parameters reducing to uncertainty in them to sub-percent level \cite{JUNOAn_2016}.

Despite the fact that three flavour neutrino oscillation framework can successfully explain most of the available oscillation data, there can be several new physics scenarios \cite{Arg_elles_2020,snow_2022} that arise as a consequence of BSM physics which may effectively influence neutrino oscillations. One such new physics scenario may result in non-standard interactions \cite{PhysRevD.17.2369,Dev_2019,Li:2014mlo} (NSI) of neutrinos with ordinary matter and it may significantly modify neutrino oscillations and possibly affect the mass hierarchy sensitivity of such experiments. 

The NSIs can be divided into two parts namely vector NSI and scalar NSI. For vector NSI, new interactions are mediated by vector bosons and can be parameterised with vector current akin to the ordinary matter effect \cite{PhysRevD.17.2369} but with unknown couplings. On the other hand, in case of scalar NSI, neutrinos can couple to the scalar fields giving rise to correction in neutrino mass term \cite{PhysRevLett.122.211801}.  The impact of vector NSI that acts like a new term to the standard matter potential has been discussed with great details in \cite{Miranda:2015dra,Farzan:2017xzy,Babu_2020,Biggio:2009nt,Ohlsson:2012kf,Agarwalla:2021zfr,Kumar:2021lrn,Farzan:2017xzy, Dev_2019, Chatterjee:2020kkm, Coloma:2015kiu, deGouvea:2015ndi, Esteban:2020cvm, Denton:2020uda, Linder:2005fc, Nieves:2003in, Nishi:2004st, Ohlsson:2012kf, Grossman:1995wx, Masud:2015xva,Masud:2018pig,Masud:2021ves}. The mass ordering sensitivities of T2K, NO$\nu$A and other long-baseline (LBL) neutrino oscillation experiments are considerably affected by the presence of vector NSI \cite{Farzan:2017xzy,Masud:2016,Capozzi:2019,Esteban:2019lfo}. In Ref. \cite{Marzec:2022mcz}, the authors have shown that the mass ordering sensitivity of JUNO may worsen if neutrino decoherence is included in the standard three flavour model. Recently, there has been a growing interest in exploring the effects of scalar NSI in neutrino oscillation experiments  \cite{PhysRevLett.122.211801, Khan:2019jvr,Smirnov:2019cae,Medhi:2021wxj,Medhi:2022qmu,Sarkar:2022ujy}. 
In \cite{PhysRevLett.122.211801}, the authors attempt to explain data from the Borexino experiment \cite{Borexino:2017rsf} with the extended paradigm where the theory includes oscillations as well as scalar NSI . Constraints on scalar NSI parameters can be obtained from big bang nucleosynthesis \cite{Venzor:2020ova}, and studies have been done on these topics using various astrophysical and cosmological considerations \cite{Babu_2020}. To date, the most stringent constraints on scalar NSI parameters are obtained from the analyses of solar neutrino data as shown in~\cite{Denton:2024upc}\footnote{The bounds from solar neutrinos are much tighter than the ones expected from LBL experiments~\cite{PhysRevLett.122.211801, Khan:2019jvr,Smirnov:2019cae,Medhi:2021wxj,Medhi:2022qmu,Sarkar:2022ujy,Denton:2022pxt} that utilize terrestrial neutrinos. This is because scalar NSI grows with density, and the Sun provides a much larger density environment compared to the Earth. However, it is still worth exploring phenomenologically the effects of scalar NSI on LBL as well as reactor neutrinos that employ relatively known and controlled neutrino sources. }. In \cite{Denton:2022pxt}, the authors discuss how to distinguish vector NSI from the scalar one in the context of DUNE \cite{DUNE_1} experiment. 

In this work, we study for the first time, the impact of scalar NSI (sNSI) on oscillation measurements at medium- and long-baseline reactor neutrino experiments - JUNO and KamLAND,  respectively. {\it We limit ourselves to assuming only one non-zero sNSI parameter: $\eta_{\rm ee}$ and assume the mass of the lightest neutrino to be $10^{-3}~\rm eV$.} We first fit the KamLAND data \cite{KamLAND:2013rgu} assuming sNSI to be present in the theory and reassess the constraints on $\theta_{12}$\footnote{Precise measurement of $\theta_{12}$ is achieved from the MSW resonance seen in solar neutrino oscillations. We focus only on the reactor experiments; a solar neutrino analysis, though very important, is beyond the scope of the present work.}, \sdm and $\eta_{\rm ee}$. Next, we calculate the sensitivity of JUNO in constraining these oscillation parameters via simulations. For both KamLAND as well as JUNO we perform spectral rates analyses assuming realistic systematic uncertainties and backgrounds - described in more detail in later sections. 

The paper is structured as follows. In section \ref{sec:NSI}, we discuss the formalism of scalar NSI and compute the individual matrix elements of effective Hamiltonian for neutrino propagation in the presence of sNSI. In section \ref{sec:analytical}, we drive the analytical expressions for modified mixing angles and mass-squared differences considering only \ee as the active sNSI parameter. Subsequently, we present the simple probability formula to understand the results. In section \ref{sec:oscprob}, we provide a brief account of electron (anti)neutrino survival probability plots for JUNO and KamLAND considering normal and inverted mass ordering for neutrinos. In Sec. \ref{sec:expt}, we provide brief description of KamLAND and JUNO experiments. In section \ref{sec:events}, we show the measured and expected events rates at KamLAND and JUNO, respectively, for the standard oscillations and in the case of oscillations with sNSI. In Secs. \ref{sec:results1} and \ref{sec:results2}, we present our results. Finally in section \ref{sec:summary}, we summarize the work and conclusions we draw.  
 \\


\section{Scalar NSI AND EFFECTIVE HAMILTONIAN}\label{sec:NSI}

Non-standard interactions for neutrinos may also proceed via coupling to a scalar field. Such interactions are mediated by scalars. In case a scalar field $\phi$ with mass $m_\phi$ that couples with neutrinos via NSI, the effective Lagrangian for such a scalar NSI for neutrinos can be written in terms of dimension six operators as
\begin{align}
    \mathcal{L}^{\rm eff~ scalar}_{\rm NSI} = \frac{y_f Y_{\alpha\beta}}{m^2_{\phi}}[\Bar{\nu_\alpha}(p_3)\nu_\beta (p_2)][\Bar{f}(p_1)f(p_4)],\label{eq:lag_scalar}  
\end{align}
where $y_f$ and $Y_{\alpha\beta}$ are the Yukawa couplings of scalar field $\phi$ with matter fermions $f\in{\{e, u, d\}}$, and neutrinos with $\phi$, respectively. The presence of Yukawa terms in the above Lagrangian in Eq. \ref{eq:lag_scalar} ensures that it can not be written as vector type currents \cite{Nieves:2003in, Nishi:2004st}. Clearly, it is no more a matter potential term that we usually observe in the case of vector-type NSI. The corresponding Dirac equation in the presence of scalar NSI can be written in the following form:

\begin{align}
    \bar{\nu}_{\beta}\Bigg[i\partial_\mu \gamma^\mu + \Bigg(\mathcal{M}_{\beta\alpha} + \dfrac{\sum_{f}N_fy_f Y_{\alpha\beta}}{m^2_{\phi}}\Bigg)\Bigg]\nu_\alpha = 0,\label{eq:dirac_eq_scalar}
\end{align}
where $N_f$ is the number density of ambient fermions $f\in{\{e, u, d\}}$ and $\mathcal{M}_{\beta\alpha}$ is the mass marix elements for neutrinos. The effective Hamiltonian for neutrino oscillations in the presence of scalar NSI takes the form \cite{Ge:2018uhz,Ge_2020,Medhi:2021wxj}
\begin{align}
    \mathcal{H}^{\rm eff}_S \approx \frac{1}{2E_\nu}\Bigg[(\mathcal{M} + \delta M) (\mathcal{M} + \delta M)^\dagger  + 2E_\nu V_{CC}\begin{pmatrix} 
1 & 0 & 0 \\
0 & 0 & 0 \\
0 & 0 & 0
\end{pmatrix}\Bigg], \label{eq:ham_s_nsi} 
\end{align}
where $\delta M \equiv\sum_{f}\frac{N_fy_f Y_{\alpha\beta}}{m^2_\phi}$ is the contribution due the scalar NSI. In the above equation $\mathcal{M} =  U_\nu\begin{pmatrix} 
m_1 & 0 & 0 \\
0 & m_2 & 0 \\
0 & 0 & m_3\end{pmatrix}U_\nu^\dagger$ is the neutrino mass matrix in flavour basis and $V_{CC} = \pm \sqrt{2}G_F N_e$ is the matter potential term due to the CC interactions where plus sign indicates the neutrino and minus sign corresponds to the antineutrino. Here, $U_\nu$ is the standard $3\times3$ Pontecorvo-Maki-Nakagawa-Sakata
(PMNS) mixing matrix \cite{Pontecorvo:1957cp, Pontecorvo:1957qd, Pontecorvo:1967fh, ParticleDataGroup:2020ssz}, and $E_\nu$ is the neutrino energy. The neutrino masses corresponding to the three neutrino mass eigen states $\nu_1, \nu_2$ and $\nu_3$ are denoted by $m_1, m_2$ and $m_3$, respectively.
From \eqs \ref{eq:dirac_eq_scalar} and \ref{eq:ham_s_nsi}, it is evident that the effect of scalar mediated NSI is to modify the neutrino mass matrix and unlike the vector NSI case which modifies the effective matter potential, the scalar NSI appears as the correction to the mass term in the Hamiltonian. Note that $\delta M$ correction to the mass term is the same for both neutrino and antineutrino. In order to study the effect of mass term correction $\delta M$ in neutrino oscillation, it is parameterised as \cite{Borexino:2017rsf, Khan:2019jvr}
\begin{equation}
\delta M \equiv \sqrt{|\Delta m_{31}^2|} \begin{pmatrix}
\eta_{ee} & \eta_{e\mu} & \eta_{e\tau} \\
\eta_{e \mu}^\star & \eta_{\mu \mu} & \eta_{\mu \tau} \\
\eta_{e \tau}^\star & \eta_{\mu \tau}^\star & \eta_{\tau\tau}
\end{pmatrix},\label{eq:delm}
\end{equation}
where $\eta_{\alpha\beta}$ represents the strength of scalar NSI. Assuming the real scalar fields, the hermiticity of the effective Hamiltonian  (Eq. \ref{eq:ham_s_nsi}) requires the diagonal elements of the scalar NSI matrix to be real while the non-diagonal parameters can be complex. \\

One of the intriguing highlights of scalar NSI is that the oscillation probabilities depend upon the absolute neutrino masses (see later). This feature is absent in the cases of the standard neutrino oscillations in matter (SI) and oscillations that include vector NSI where oscillation probabilities are only sensitive to the mass-squared differences. Consequently, the scalar NSI model opens up possibilities to address the absolute neutrino masses from neutrino oscillation studies. Moreover, this model is independent of the choice of neutrino energy and hence can be probed in oscillation experiments of all energy ranges. However, ambient electron density i.e. earth matter is required for non-zero scalar NSI. In order to calculate mass-matrix terms free from sNSI parameters, one requires neutrino oscillation experiments performed in vacuum. Unfortunately, no such data is available and therefore, oscillation measurements done in terrestrial settings always calculate ``effective paremeters" that include sNSI effects - should sNSI exist in nature. In a previous work, \cite{PhysRevLett.122.211801}, authors have discussed the approach of density subtraction where they consider the reactor measurements as benchmark for least sNSI effects - a near-vacuum measurement. Measurements made at longer baselines with greater matter effects are to be compared against the reactor experiments to calculate an estimate of sNSI. {\it In this work, our approach is different - we start with the assumption that sNSI do exist in nature. So our model has sNSI. We then fit the KamLAND data with this sNSI model and calculate the ability of the JUNO experiment to constrain oscillation parameters in the presence of sNSI. }
%

Next, we compute the expressions for the effective Hamiltonian considering only diagonal scalar NSI parameters\footnote{It is worthwhile mentioning here that although for the analytical computations, we ignore the standard matter potential (which is a good approximation for JUNO baseline), the numerical simulations and the results shown in later sections do include the standard matter potential term.}. The effective Hamiltonian in the flavour basis including the scalar NSI can be written from \eq \ref{eq:ham_s_nsi} as
\begin{equation}
    \mathcal{H}^{\rm eff}_S \approx \frac{1}{2E_\nu}\Bigg[(\mathcal{M} + \delta M) (\mathcal{M} + \delta M)^\dagger\Bigg]. \label{eq:ham_s_nsi_vac} 
\end{equation}
 The expression for $\delta M$ in the above equation is given in Eq. \ref{eq:delm}. We can write the neutrino mass matrix \mass in flavour basis as
 \begin{eqnarray}
     \mathcal{M} &= U_\nu D_\nu {U_\nu}^\dagger = \begin{pmatrix} 
\mathcal{M}_{11} & \mathcal{M}_{12} & \mathcal{M}_{13} \\
\mathcal{M}_{21} & \mathcal{M}_{22} & \mathcal{M}_{23} \\
\mathcal{M}_{31} & \mathcal{M}_{32} & \mathcal{M}_{33}  
\end{pmatrix} = \mathcal{M}^\dagger
\label{eq:mass}
 \end{eqnarray}
 where 
 \begin{align}
     D_\nu = \begin{pmatrix} 
m_1 & 0 & 0 \\
0 & m_2 & 0 \\
0 & 0 & m_3
\end{pmatrix}, \nonumber
 \end{align}
 with 
  \begin{align}
     \mathcal{M}_{11} &= c_{12}^2c_{13}^2 m_1 + c_{13}^2s_{12}^2m_2+m_3s_{13}^2 \nonumber\\
     \mathcal{M}_{12} & = -c_{13}\big[(m_1-m_2)s_{12}c_{12}c_{23}+s_{13}s_{23}e^{-i\delta_{CP}}(c_{12}^2m_1 - m_3 + s_{12}^2m_2)\big]\nonumber\\
     \mathcal{M}_{13} &= -c_{13}\big[(m_2-m_1)s_{12}c_{12}s_{23}+s_{13}c_{23}e^{-i\delta_{CP}}(c_{12}^2m_1-m_3+s_{12}^2m_2)\big]\nonumber\\
     \mathcal{M}_{21} &= \mathcal{M}_{12}^\star\nonumber\\
     \mathcal{M}_{22} & = m_1s_{12}^2c_{23}^2+m_3s_{23}^2c_{13}^2+m_2s_{12}^2s_{13}^2s_{23}^2 + c_{12}^2(m_2c_{23}^2+m_1s_{13}^2s_{23}^2) \nonumber\\
     &+2s_{12}c_{12}s_{23}c_{23}s_{13}(m_1-m_2)\cos{\delta_{CP}}\nonumber\\
     \mathcal{M}_{23}& = s_{23}c_{23}\big[m_3c_{13}^2+c_{12}^2(-m_2+m_1s_{13}^2) + s_{12}^2(-m_1+m_2s_{13}^2)\big]\nonumber\\
     &+c_{12}(m_1-m_2)s_{13}^2(c_{23}^2e^{-i\delta{CP}}-s_{23}^2e^{i\delta_{CP}})\nonumber\\
     \mathcal{M}_{31}& = \mathcal{M}_{13}^\star\nonumber\\
     \mathcal{M}_{32} & = \mathcal{M}_{23}^\star\nonumber\\
     \mathcal{M}_{33}& = m_3c_{13}^2c_{23}^2+m_1c_{12}^2c_{23}^2s_{13}^2 + m_2s_{12}^2s_{13}^2c_{23}^2 + m_2s_{23}^2c_{12}^2+m_1s_{12}^2s_{23}^2\nonumber\\
     &-2s_{12}c_{12}s_{23}c_{23}s_{13}(m_1-m_2)\cos{\delta_{CP}}\,.\label{eq:element m}
 \end{align}
In the above, $s_{ij} = \sin{\theta_{ij}}$ and $c_{ij} = \cos{\theta_{ij}}$, and $*$ denotes the complex conjugate. 

The simplified form of the effective Hamiltonian including scalar NSI is obtained by neglecting the term $\delta M^2$ in Eq. \ref{eq:ham_s_nsi}, which is $\sim$ $1000$ times small compared to other terms and we approximate \eq \ref{eq:ham_s_nsi} as 
\begin{align}
    \mathcal{H}^{\rm eff}_S & \approx \dfrac{\Delta m_{31}^2}{2E_\nu} \left[U_\nu\begin{pmatrix}
0 & 0 & 0 \\
0& \alpha& 0 \\
0& 0 & 1
\end{pmatrix}{U_\nu}^\dagger + \dfrac{1}{\Delta m_{31}^2} (\mathcal{M}\delta M^\dagger + \delta M \mathcal{M}^\dagger)\right] , \label{eq:hf_simple}
\end{align}
where $\alpha = \dfrac{\Delta m_{21}^2}{\Delta m_{31}^2}$. The first term is responsible for the standard oscillation whereas the second one is due to the scalar NSI. If we define $\widetilde{\mathcal{H}} = \dfrac{1}{\Delta m^2_{31}} [$\mass$\delta M^\dagger +  \delta M \mathcal{M}^\dagger ] = \begin{pmatrix}
h_{11} & h_{12} & h_{13} \\
h_{12}^\star& h_{22}& h_{23} \\
h_{13}^\star& h_{23}^\star & h_{33}
\end{pmatrix}$, then one can rewrite the elements of effective matrix elements as  
\begin{align}
    {\mathcal{H}^{\rm eff}_S}^{11} &= \dfrac{\Delta m_{31}^2}{2E_\nu} (s_{13}^2+\alpha s_{12}^2c_{13}^2+h_{11})\nonumber\\
    {\mathcal{H}^{\rm eff}_S}^{12} &= \dfrac{\Delta m_{31}^2}{4E_\nu}\big[\sin{2\theta_{13}}(1-\alpha s_{12}^2)s_{23}e^{-i\delta_{CP}} +\alpha c_{13}c_{23}\sin{2\theta_{12}}+2h_{12}\big]\nonumber\\
    {\mathcal{H}^{\rm eff}_S}^{13} &= \dfrac{\Delta m_{31}^2}{4E_\nu}\big[\sin{2\theta_{13}}(1-\alpha s_{12}^2)c_{23}e^{-i\delta_{CP}}-\alpha \sin{2\theta_{12}}c_{13}s_{23}+2h_{13}\big]\nonumber\\
    {\mathcal{H}^{\rm eff}_S}^{22} &= \dfrac{\Delta m_{31}^2}{2E_\nu} \big[ c_{13}^2s_{23}^2+\alpha c_{12}^2c_{23}^2+\alpha s_{12}^2s_{13}^2s_{23}^2-\alpha s_{12}c_{12}s_{23}c_{23}s_{13}(e^{-i\delta_{CP}}+e^{i\delta_{CP}})+ h_{22}\big]\nonumber\\
    {\mathcal{H}^{\rm eff}_S}^{23} &= \dfrac{\Delta m_{31}^2}{4E_\nu}\big[\sin{2\theta_{23}}c_{13}^2 +\alpha \sin{2\theta_{23}}(s_{12}^2s_{13}^2-c_{12}^2)+\alpha\sin{2\theta_{12}}s_{13}(s_{23}^2e^{i\delta_{CP}}-c_{23}^2e^{-i\delta_{CP}})+2h_{23}\big]\nonumber\\
    {\mathcal{H}^{\rm eff}_S}^{33} &= \dfrac{\Delta m_{31}^2}{2E_\nu}\big[ c_{13}^2c_{23}^2+\alpha c_{12}^2s_{23}^2+\alpha s_{12}^2s_{13}^2c_{23}^2+\alpha s_{12}c_{12}s_{23}c_{23}s_{13}(e^{-i\delta_{CP}}+e^{i\delta_{CP}})+ h_{33}\big], \label{eq:Hf_element}
\end{align}
with 
\begin{align}
    h_{11} & = 2r\eta_{ee} \big[c_{13}^2(m_1c_{12}^2+m_2s_{12}^2) + m_3s_{13}^2\big]\nonumber\\
    h_{12} & = c_{13}r(\eta_{ee}+\eta_{\mu\mu})\big[-s_{12}c_{12}c_{23}d + s_{13}s_{23}e^{-i\delta_{CP}}(-m_1c_{12}^2+m_3-m_2s_{12}^2)\big]\nonumber\\
    h_{13} & = c_{13}r(\eta_{ee}+\eta_{\tau\tau}\big[s_{12}c_{12}s_{23}d + s_{13}c_{23}e^{-i\delta_{CP}}(-m_1c_{12}^2+m_3-m_2s_{12}^2)\big]\nonumber\\
    h_{22}& = 2r\eta_{\mu\mu} \big[c_{23}^2(m_1s_{12}^2+m_2c_{12}^2)+s_{23}^2\big(m_3c_{13}^2+s_{13}^2(m_1c_{12}^2+m_2s_{12}^2)\big) + d \sin{2\theta_{12}}s_{23}c_{23}s_{13}\cos{\delta_{CP}}\big]\nonumber\\
    h_{23}&= r e^{-i\delta_{CP}}(\eta_{\mu\mu}+\eta_{\tau\tau})\big[ds_{12}c_{12}c_{23}^2s_{13}+ s_{23}c_{23}e^{i\delta_{CP}}\big(m_3c_{13}^2+c_{12}^2(m_1s_{13}^2-m_2) +s_{12}^2(m_2s_{13}^2-m_1)\big)\nonumber\\
    &-d c_{12}s_{12}s_{13}s_{23}^2e^{2i\delta_{CP}}\big]\nonumber\\
    h_{33}& = 2r\eta_{\tau\tau}\big[s_{23}^2(m_1s_{12}^2+m_2c_{12}^2)+c_{23}^2\big(m_3c_{13}^2+s_{13}^2(m_1c_{12}^2+m_2s_{12}^2)\big) - d \sin{2\theta_{12}}s_{23}c_{23}s_{13}\cos{\delta_{CP}}\big]. \label{eq:hij}
\end{align}
Here $r = \dfrac{\sqrt{|\Delta m^2_{31|}}}{\Delta m^2_{31}}$ and $d = m_1 - m_2$. One can easily observe from Eqs. \ref{eq:Hf_element} and \ref{eq:hij} that all the elements of effective Hamiltonian explicitly depend on the individual neutrino masses owing to the presence of sNSI.


\section{Analytical Expressions for Electron (Anti)Neutrino Survival Probability}
\label{sec:analytical}

In this section, the formula for $\bar{\nu}_e\rightarrow\bar{\nu}_e$ disappearance probability ($P_{\bar{\nu}_e\to \bar{\nu}_e}$) in the presence of scalar NSI is derived. As mentioned earlier, we consider the diagonal scalar NSI parameters only for simplicity. Moreover, one sNSI parameter at a time is taken while the other two NSI parameters are put to zero. Thus for the derivation of $P_{\bar{\nu}_e\to \bar{\nu}_e}$
(with sNSI) only the sNSI parameter $\eta_{ee}$ takes non-zero values while $\eta_{\mu\mu}$ $= 0 = $ \tt. Here we use the change of basis approach \cite{Ioannisian:2018qwl} to diagonalize the effective Hamiltonian in the presence of sNSI.  

The neutrino evolution equation in flavour basis is given by
\begin{align}
    i\dfrac{\partial}{\partial t}\ket{\nu_f}= \mathcal{H}^{\rm eff}_S \ket{\nu_f}\,,
    \label{eq:evol1}
\end{align}
where the effective neutrino Hamiltonian $\mathcal{H}^{\rm eff}_S$ is given in \eq \ref{eq:ham_s_nsi_vac} and can be rewritten as
\begin{align}
   \mathcal{H}^{\rm eff}_S \approx  \dfrac{\left(U_\nu D_\nu U_\nu^\dagger +\delta M\right)\left(U_\nu D_\nu U_\nu^\dagger+\delta M\right)^\dagger}{2E_\nu}\,.  
   \label{eq:Heff2}
\end{align}

Neglecting the term proportional to $\delta M^2$ in \eq \ref{eq:Heff2}, we can write the evolution equation  \ref{eq:evol1} with the terms relevant for neutrino oscillation as

\begin{align}
    i\dfrac{\partial}{\partial t}\ket{\nu_f} & = \dfrac{1}{2E_\nu}\Big[U_\nu D^2_\nu U^\dagger_\nu + U_\nu D_\nu U^\dagger_\nu(\delta M) +(\delta M)U_\nu D_\nu U^\dagger_\nu\Big]\ket{\nu_f}\nonumber\\
   &= \dfrac{1}{2E_\nu}\Big[U_\nu M^2_d U^\dagger_\nu +  \mathcal{M}(\delta M) +(\delta M) \mathcal{M}\Big]\ket{\nu_f}
    \,.  \label{eq:evol2}
\end{align}
In the above, 
\begin{align}
   M^2_d = \begin{pmatrix}
0 & 0 & 0 \\
0 & \Delta m^2_{21} & 0 \\
0 & 0 & \Delta m^2_{31}\end{pmatrix} =\begin{pmatrix}
0 & 0 & 0 \\
0 & \alpha \Delta m^2_{31} & 0 \\
0 & 0 & \Delta m^2_{31}\end{pmatrix}\,.\nonumber
\end{align}
 Defining $U_\nu \left(=U_{23}U_\delta U_{13} U_\delta^\dagger U_{12}\right) = U_A U_B$ where $U_A = U_{23} U_\delta$ and $U_B = U_{13}U^\dagger_\delta U_{12}$, \eq \ref{eq:evol2} takes the form
\begin{align}
    i\dfrac{\partial}{\partial t}\ket{\nu_f} &= \dfrac{1}{2E_\nu}\Big[U_A U_B M^2_d U^\dagger_B U^\dagger_A + \mathcal{M}(\delta M) +(\delta M) \mathcal{M}\Big]\ket{\nu_f}\nonumber\\
    & = \dfrac{1}{2E_\nu}\Big[U_A U_B M^2_d U^\dagger_B U^\dagger_A + U_A U_B D_\nu U^\dagger_B U^\dagger_A(\delta M) +(\delta M) U_A U_B D_\nu U^\dagger_B U^\dagger_A\Big]\ket{\nu_f}\nonumber\\
    & = \dfrac{1}{2E_\nu}U_A\Big[ U_B M^2_d U^\dagger_B + U_B D_\nu U^\dagger_B(\delta M) +(\delta M)  U_B D_\nu U^\dagger_B \Big]U^\dagger_A\ket{\nu_f}\,. \label{eq:evol3}
\end{align}
With the assumption that only \ee is non-zero for the present case, $\delta M \equiv \sqrt{|\Delta m_{31}^2|} \begin{pmatrix}
\eta_{ee} & 0 & 0 \\
0 & 0 & 0 \\
0 & 0 & 0
\end{pmatrix}$. In Eq. \ref{eq:evol3}, we have used the fact that this form of $\delta M$ allows us to write $(\delta M) U_A = U_A (\delta M)$ and $(\delta M) U^\dagger_A = U^\dagger_A (\delta M)$. 

Defining a new basis $\ket{\widetilde{\nu_f}} =U^\dagger_A\ket{\nu_f}$ and identifying that $[\delta M, U_A] = 0$,  \eq \ref{eq:evol3} modifies as  
\begin{align}
 i\dfrac{\partial}{\partial t}2E_\nu \ket{\widetilde{\nu_f}} =M^{\rm eff}\ket{\widetilde{\nu_f}} = \Big[M_0 + M_1\Big]\ket{\widetilde{\nu_f}}\label{eq:evol4}    
\end{align}
with $M_0 = U_B M^2_d U^\dagger_B$ and $M_1 = U_B D_\nu U^\dagger_B (\delta M) + (\delta M)U_BD_\nu U^\dagger_B$.

The eigenvalues of $M^{\rm eff}$ as also the modified mixing matrix $U^\prime_\nu$ are derived by first choosing an auxiliary basis $U^{\rm aux} = U_{13}$ such that 
\begin{align}
M^\prime &= {U^{\rm aux}}^\dagger M^{\rm eff} U^{\rm aux}\nonumber\\
    &= \begin{pmatrix}
\alpha s^2_{12} \Delta m^2_{31} +\epsilon_2& \alpha\Delta m^2_{31}s_{12}c_{12}+ s_{12}c_{12}c^2_{13}\epsilon_3 & s_{31}c_{13}\epsilon_1 \\
\alpha\Delta m^2_{31}s_{12}c_{12}+ s_{12}c_{12}c^2_{13}\epsilon_3 & \alpha\Delta m^2_{31}c^2_{12} & s_{12}c_{12}s_{3}c_{13}\epsilon_3 \\
 s_{31}c_{13}\epsilon_1 & s_{12}c_{12}s_{3}c_{13}\epsilon_3 & \Delta m^2_{31}+2m_3\beta\eta_{ee}s^2_{13}
\end{pmatrix}.  
\end{align}
\\
Here, $\epsilon_1 = \beta\eta_{ee}(m_1c^2_{12}+m_2s^2_{12}+m_3)$, $\epsilon_2 = 2\beta\eta_{ee}c^2_{13}(m_1c^2_{12}+ m_2s^2_{12})$, and $\epsilon_3= \beta\eta_{ee}(-m_1+m_2)$ and $\beta = \sqrt{\vert\Delta m^2_{31}\vert}$. \\
 We then apply two successive rotations $O^\prime_{13} $ $(1-3$ rotation) followed by $O^\prime_{12}$ ($1-2$ rotation) and finally obtain the modified mixing angles by substituting the off-diagonal elements of each $2\times2$ sub-matrix equal to zero after every rotation \cite{Ioannisian:2018qwl}. We have found that after two rotations $M^{\rm eff}$ is diagonal. The modified mixing angles are given by 
\begin{equation}
   \tan{2\theta_{13}}^\prime = \dfrac{\sin{2\theta_{13}}\epsilon_1}{\Delta m^2_{31}(1-\alpha s^2_{12}) - \epsilon_2}\label{eq:th13} ~~\text{and} 
\end{equation}
\begin{align}
    \tan{2\theta_{12}}^\prime =  \dfrac{-\sin{2\theta_{12}}c_{13}(c_{13}\cos{\theta}\epsilon_3 + \alpha\Delta m^2_{31}c^\prime_{13})}{c_{13}\Delta m^2_{31}({s^\prime_{13}}^2 + \alpha s^2_{12}{c^\prime_{13}}^2) -\alpha\Delta m^2_{31}c_{13}c^2_{12} + c^\prime_{13}\epsilon_2\cos{\theta}}\label{eq:th12}
\end{align}
where, $\cos{\theta} = c_{13}c^\prime_{13} - s_{13}s^\prime_{13}$. The corresponding eigenvalues of the $M^{\rm eff}$ are computed as
\begin{align}
    \Lambda_1 &=  \Delta m^2_{31}(s^\prime_{13}c^\prime_{12})^2 + \alpha\Delta m^2_{31}(c^\prime_{12}c^\prime_{13}s_{12}-c_{12}s^\prime_{12})^2 \nonumber \\
    & + 2\beta\eta_{ee}\cos{\theta}c^\prime_{12}\Big[c^\prime_{12}c_{13}c^\prime_{13}(m_1c^2_{12}+m_2s^2_{12})+ s_{12}c_{12}c_{13}s^\prime_{12}(m_1-m_2)\Big]\label{eq:eval1}\\
    \Lambda_2 & = \Delta m^2_{31}(s^\prime_{12} s^\prime_{13})^2 + \alpha\Delta m^2_{31}(c_{12}c^\prime_{12} + s_{12}s^\prime_{12}c^\prime_{13})^2\nonumber\\
    &+ 2\beta\eta_{ee}\cos{\theta}s^\prime_{12}c_{31}\Big[s_{12}c_{12}c^\prime_{12}(-m_1+m_2)+ s^\prime_{12}c^\prime_{13}(m_1c^2_{12}+m_2s^2_{12}) \Big]\label{eq:eval2}\\
    \Lambda_3 & = \Delta m^2_{31} {c^\prime_{13}}^2 + 2\beta\eta_{ee}\sin{\theta}\Big[m_3s_{13}c^\prime_{13}+c_{13}s^\prime_{13}(m_1c^2_{12}+m_2s^2_{12})\Big],\label{eq:eval3}
\end{align}
where, $\sin\theta = s_{13}c^\prime_{13} + c_{13}s^\prime_{13}$. Finally, we obtain the PMNS mixing matrix in the presence of scalar NSI, $\eta_{ee}$ as
\begin{align}
    U^\prime_\nu &= U_A U^{\rm aux} O^\prime_{13}O^\prime_{12}\nonumber\\
    & = U_{23}(\theta_{23})U_\delta U_{13}(\theta_{13})O^\prime_{13}(\theta^\prime_{13})O^\prime_{12}(\theta^\prime_{12})\nonumber\\
    & =U_{23}(\theta_{23})U_\delta \widetilde{U}_{13}(\widetilde{\theta}_{13})O^\prime_{12}(\theta^\prime_{12})\\
    & = \begin{pmatrix}
c^\prime_{12}\Tilde{c}_{13} & s^\prime_{12}\Tilde{c}_{13} & \Tilde{s}_{13} \\
-c_{23}s^\prime_{12}-c^\prime_{12}\Tilde{s}_{13}s_{23}e^{i\delta} & c_{23}c^\prime_{12}-s^\prime_{12}\Tilde{s}_{13}s_{23}e^{i\delta} & \Tilde{c}_{13}s_{23}e^{i\delta} \\
s_{23}s^\prime_{12}-c^\prime_{12}\Tilde{s}_{13}c_{23}e^{i\delta} & -s_{23}c^\prime_{12}-s^\prime_{12}\Tilde{s}_{13}c_{23}e^{i\delta} & \Tilde{c}_{13}c_{23}e^{i\delta}\end{pmatrix}
    \label{eq:umix}
\end{align}
where, $\Tilde{s}_{13} = \sin{\Tilde{\theta}_{13}}$ and $\widetilde{\theta}_{13} = \theta_{13}+\theta^\prime_{13}$. Thus, scalar NSI modified mixing matrix is easy to obtain from the vacuum mixing matrix by just replacing the following parameters in the vacuum PMNS matrix as
\begin{align}
   \theta_{23} &\to  \theta^\prime_{23} (=\theta_{23})\nonumber\\
    & \delta\to \delta^\prime (=\delta)  \nonumber\\
    & \theta_{13} \to \Tilde{{\theta}}_{13}\nonumber\\
    & \theta_{12}\to \theta^\prime_{12}\,.\label{eq:change}
\end{align} 
The electron antineutrino ($\bar{\nu}_e$) survival probability in vacuum (i.e. without including sNSI) for JUNO baselines is given by \cite{Ge:2012wj}
 
 \begin{eqnarray} 
 P_{\bar{\nu}_e\to\bar{\nu}_e} = P_{ee} = 1 &-& \cos^4{\theta_{13}}\sin^2{2\theta_{12}}\sin^2{\Delta_{21}} \nonumber \\
        &-& \sin^22\theta_{13}\sin^2\left(\vert\Delta_{31}\vert\right) \nonumber \\ 
        &-& \sin^2\theta_{12}\sin^22\theta_{13}\sin^2{\Delta_{21}}\cos\left(2\vert\Delta_{31}\vert\right) \nonumber \\ 
        &\pm& \frac{\sin^2\theta_{12}}{2}\sin^22\theta_{13}\sin\left(2\Delta_{31}\right)\sin\left(2\vert\Delta_{31}\vert\right). 
        \label{eq:pee_ana}
 \end{eqnarray} 
where $\Delta_{ij}\equiv \dfrac{\Delta m^2_{ij}L}{4E_\nu} $. While for  KamLAND baselines, we have the approximate probability expression given by 
\cite{KamLAND:2013rgu}
\begin{equation} 
P_{ee} = \cos^4\theta_{13} (1-\sin^22\tilde{\theta}_{12}\sin^2(\Delta \tilde{m}^2_{21}L/4E)) + \sin^4\theta_{13}.
\label{eq:pee_kamland}
\end{equation} 
where $\tilde{\theta}_{12}$ and $\Delta \tilde{m}^2_{21}$ are matter-modified mixing angle and mass-squared difference, respectively. 

We calculate the value of $P_{ee}$ analytically in the presence of $\eta_{ee}$ by substituting in \eq \ref{eq:pee_ana} the eigenvalues from vacuum to the sNSI modified mass eigenvalues obtained in \eqs \ref{eq:eval1}, \ref{eq:eval2}, \ref{eq:eval3} and substituting the mixing angles given by \eq \ref{eq:change}. The results match very well with the numerically-obtained exact probability values for $P(\bar{\nu}_{e}\rightarrow \bar{\nu}_{e})$ and also other oscillation channels. In the above equation, only the last term is sensitive to the mass ordering where the plus sign corresponds to NO and the minus corresponds to IO. We also note that the dependence on $\theta_{13}$-terms is only via $\sin^22\theta_{13}$ and therefore is a second-order effect.

We further simplify the correction terms in the expressions for modified mixing angles and mass-squared differences to understand the effects of $\eta_{ee}$ in the case of NO and IO. From \eq \ref{eq:th13}, we note that the $\eta_{ee}$-correction to $\theta_{13}$ is itself $\theta_{13}$-suppressed. \textit{Thus, we expect $\theta_{13}$ measurements not to be affected by $\eta_{ee}$.} Further, since in \eq \ref{eq:pee_ana}, $\theta_{13}$-terms are of second-order, we do not expect any significant effect of $\eta_{ee}$ on $P_{ee}$ through $\theta_{13}^\prime$. In order to assess the effect of $\eta_{ee}$ on the dominant $\theta_{12}$ term, we approximate $\theta_{13}^\prime \approx 0$. Thus, with $\theta_{13}^\prime = 0$ in \eq\ref{eq:th12} we get
\begin{align}
    \tan{2\theta_{12}}^\prime =  \dfrac{\sin2\theta_{12}\left(\Delta m^2_{21} + \epsilon_{3}\cos^2{\theta_{13}}\right)}{\Delta m^2_{21}\cos2\theta_{12} - \epsilon_{2}}.\label{eq:th12-1}
\end{align} 
Therefore, for $\epsilon_{2},\epsilon_{3}\ll \Delta m^2_{21}$, we do not expect any significant modification in $\theta_{12}$ because of $\eta_{ee}$. Similarly, for the mass-squared differences we have 
 \begin{eqnarray} 
 {\Delta m^2_{21}}^\prime &=& \Lambda_{2} - \Lambda_{1} \nonumber \\
        &=& \Delta m^2_{21}\cos2\left(\theta_{12} - \theta_{12}^\prime\right) + \cos^2\theta_{13}\sin2\theta_{12}\sin2\theta_{12}^\prime\epsilon_{3} - \cos2\theta_{12}^\prime\epsilon_{2}\,.
        \label{eq:dm21nsi}
 \end{eqnarray} 
The $\eta_{ee}$-correction to $\Lambda_{3}$ is also suppressed by $\sin\theta_{13}$. On substituting $\theta_{13}^\prime=0$, we get $\Lambda_{3} = \Delta m^2_{31} + 2\beta\eta_{ee}m_{3}\sin^2\theta_{13}$. Hence, we approximate $\Lambda_3 \approx \Delta m^2_{31}$. So $\eta_{ee}$-modified atmospheric mass-squared difference is
 \begin{eqnarray} 
{\Delta m^2_{31}}^\prime &=& \Lambda_{3} - \Lambda_{1} \nonumber \\        
        &=& \Delta m^2_{31} - \Delta m^2_{21}\sin^2\left(\theta_{12} - \theta^\prime_{12}\right)  \nonumber \\ && - {\cos^2\theta^\prime_{12}}\epsilon_{2} - \frac{1}{2}\cos^2\theta_{13}\sin2\theta_{12}\sin2\theta_{12}^\prime\epsilon_{3}\,.
        \label{eq:dm31nsi}
 \end{eqnarray} 
The information about sNSI $\eta_{ee}$ is encoded in the $\epsilon$ parameters. The $\epsilon$'s can be expressed in terms of the ratios of mass eigenvalues. For NO, we have $\epsilon_1 = \beta\eta_{ee}m_1(c^2_{12}+\frac{m_2}{m_1}s^2_{12}+\frac{m_3}{m_1})$, $\epsilon_2 = 2\beta\eta_{ee}c^2_{13}m_1(c^2_{12}+ \frac{m_2}{m_1}s^2_{12})$, and $\epsilon_3= \beta\eta_{ee}m_1(-1+\frac{m_2}{m_1})$.  
For IO, $\epsilon_1 = \beta\eta_{ee}m_3(\frac{m_1}{m_3}c^2_{12}+\frac{m_2}{m_3}s^2_{12}+1)$, $\epsilon_2 = 2\beta\eta_{ee}c^2_{13}m_3(\frac{m_1}{m_3}c^2_{12}+ \frac{m_2}{m_3}s^2_{12})$, and $\epsilon_3= \beta\eta_{ee}m_3(-\frac{m_1}{m_3}+\frac{m_2}{m_3})$. Assuming lightest neutrino mass ($m_1$ for NO and $m_3$ for IO) $\sim 0.001$ eV, we have $\frac{m_2}{m_1}\approx8$ and $\frac{m_3}{m_1}\approx50$ for NO. For IO, both $\frac{m_1}{m_3}$ and $\frac{m_2}{m_3}$ $\approx50$. For NO, $m_2 = \sqrt{m^2_{1}+ \Delta m^2_{21}}$ and $m_3 = \sqrt{m^2_{1} + \Delta m^2_{31}}$ while for IO, $m_1 = \sqrt{m_3^2 + |\Delta m^2_{31}|}$ and $m_2 = \sqrt{m^2_3 + |\Delta m^2_{31}| + \Delta m^2_{21}}$. To understand how $\eta_{ee}$ affects the oscillation probability P$_{ee}$ differently for NO and IO cases, we approximate Eq. \ref{eq:pee_ana} only up to the leading order terms neglecting the terms suppressed by higher orders of $\sin{\theta_{13}}$.  
\begin{eqnarray} 
P_{ee} \approx 1 &-& \cos^4{\theta_{13}}\sin^2{2\theta_{12}}\sin^2{\Delta_{21}}.
        \label{eq:pee_approx}
 \end{eqnarray} 
The modified value of $\theta_{12}$ is given by Eq.~\ref{eq:th12-1}. The parameters $\epsilon_2$ and $\epsilon_3$ which are related to $\eta_{ee}$ should be comparable with \sdm in order to have significant corrections in $\theta_{12}$. These parameters behave differently for IO and NO cases. First we notice that for any non-zero value of $\eta_{ee}$,  $\epsilon_2(\rm IO) \approx 16 \epsilon_2(\rm NO)$ and $\epsilon_3(\rm IO) \approx \frac{1}{11} \epsilon_3(\rm NO)$. Moreover, For NO, $\epsilon_2 (\rm NO) \sim \epsilon_3(\rm NO)$  whereas for IO,    $\epsilon_2 (\rm IO) \sim 100\epsilon_3(\rm IO)$. (For instance, if we consider $\eta_{\rm ee} = 0.04, \epsilon_2(\rm IO) = 2.26\times 10^{-4}; \epsilon_3(\rm IO) = 1.47\times 10^{-6}; \epsilon_2(\rm NO) = 1.37\times 10^{-5}; \epsilon_3(\rm NO) = 1.53\times 10^{-5}$. As a result, the relative change in $\sin^22\theta_{12}$ compared to its vacuum value is $\sim$ 0.35 for NO and 0.85 for IO.) Thus, due to the larger value of $\epsilon_2(\rm IO)$ compared to   $\epsilon_2(\rm NO)$ for any given $\eta_{\rm ee}$ the value of $\sin^22\theta_{12}$ enhances substantially for the case of IO. This increases the amplitude of the oscillation probability P$_{ee}$ more in the case of IO than NO. A similar argument can be made for the correction to mass-squared difference \sdm.  


\begin{center}
\begin{table}
\begin{tabular}{|c |c| c|} 
 \hline
Oscillation parameters ($3\nu$) & Normal ordering (NO) & Inverse Ordering (IO) \\ [0.5ex] 
 \hline\hline
$\theta_{12}^{\circ}$ & $33.41^{+0.75}_{-0.72}$ & $33.41^{+0.75}_{-0.72}$\\
 \hline
$\theta_{23}^{\circ}$ & $42.2^{+1.1}_{-0.9}$ &  $49.0^{+1.0}_{-1.2}$\\
\hline
$\theta_{13}^
{\circ}$ & $8.58^{+0.11}_{-0.11}$ &  $8.57^{+0.11}_{-0.11}$\\
\hline
$\delta_{CP}^{\circ}$ & $232^{+36}_{-26}$ & $276^{+22}_{-29}$ \\
\hline
$\Delta m_{21}^2$ (eV$^2$) & $7.41^{+0.21}_{-0.20}\times 10^{-5}$ & $7.41^{+0.21}_{-0.20}\times 10^{-5}$\\
\hline
$\Delta m_{31}^2$ (eV$^2$) & $+2.507^{+0.026}_{-0.027}\times 10^{-3}$ & $-2.486^{+0.025}_{-0.028}\times 10^{-3}$\\
\hline
\end{tabular}
\caption{Best-fit values of the neutrino oscillation parameters  considering the standard three-flavour scenario. These values with $1\sigma$ interval are taken from NuFIT 5.2 (2022), including Super-K atmospheric data \cite{Esteban:2020cvm}}.    \label{tab:t1}
\end{table}
\end{center}


\section{Oscillation probability in the presence of scalar NSI} 
\label{sec:oscprob} 

\begin{figure}[h]
\centering
\includegraphics[width=.49\textwidth]{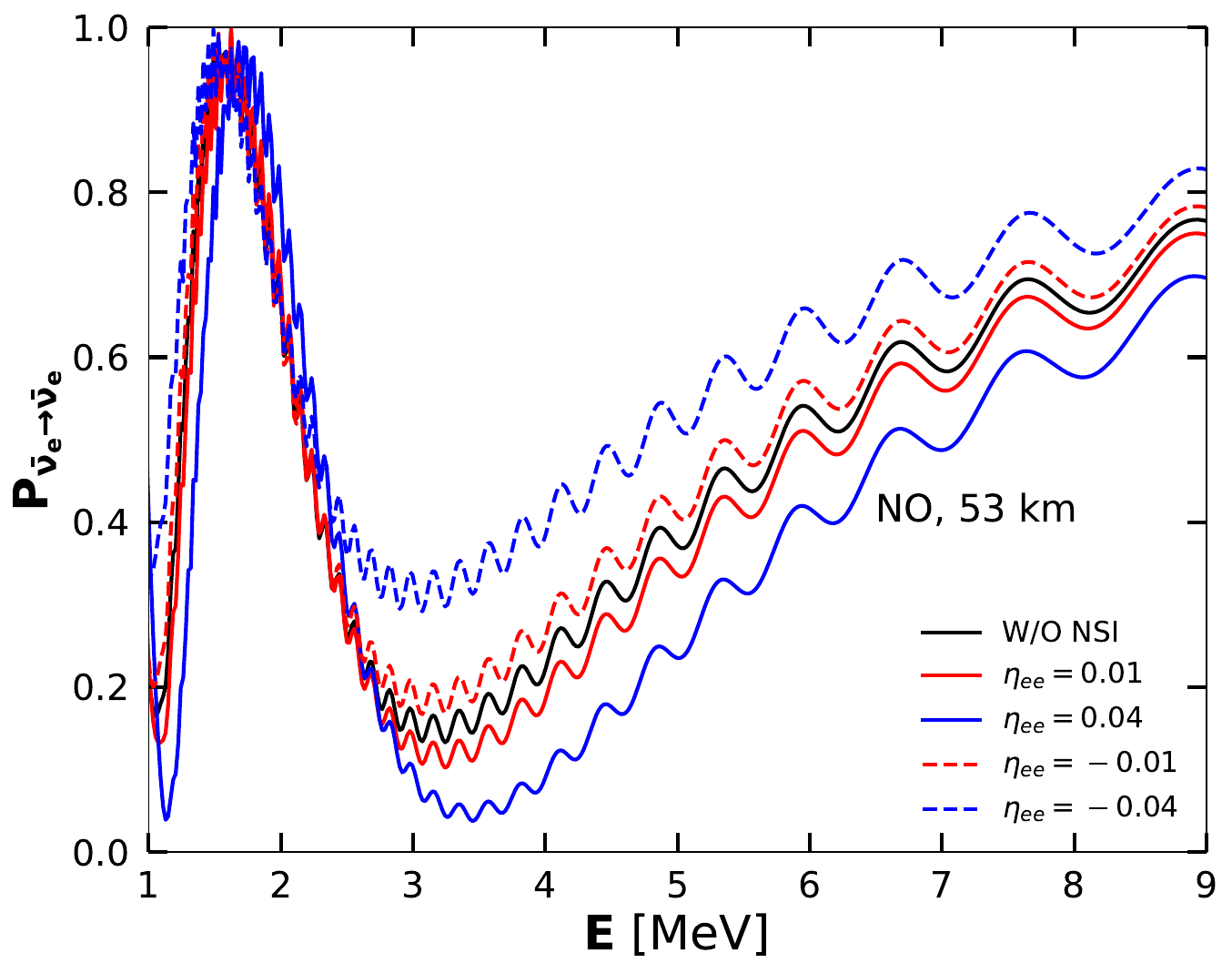}
\includegraphics[width=.49\textwidth]{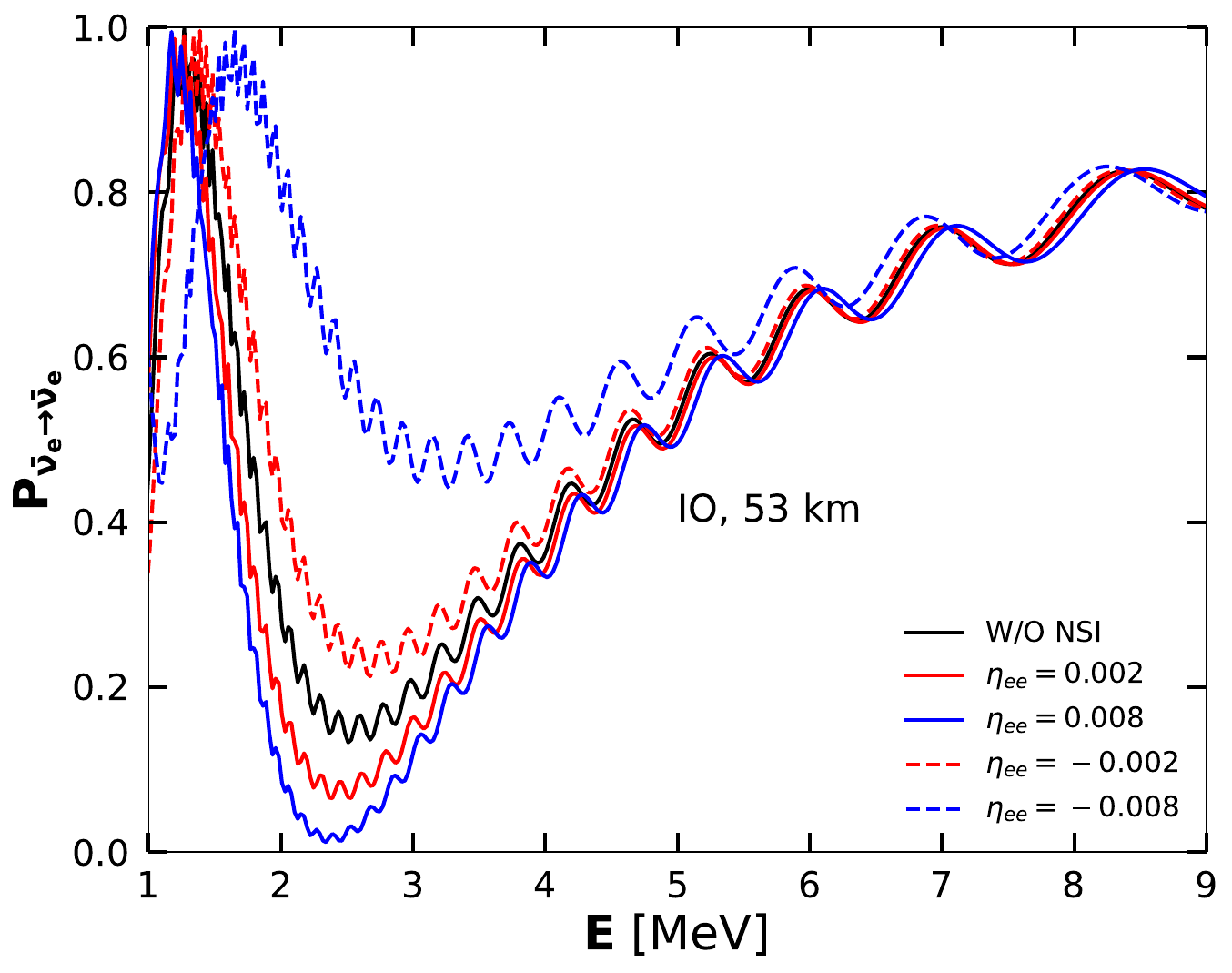}
\includegraphics[width=.49\textwidth]{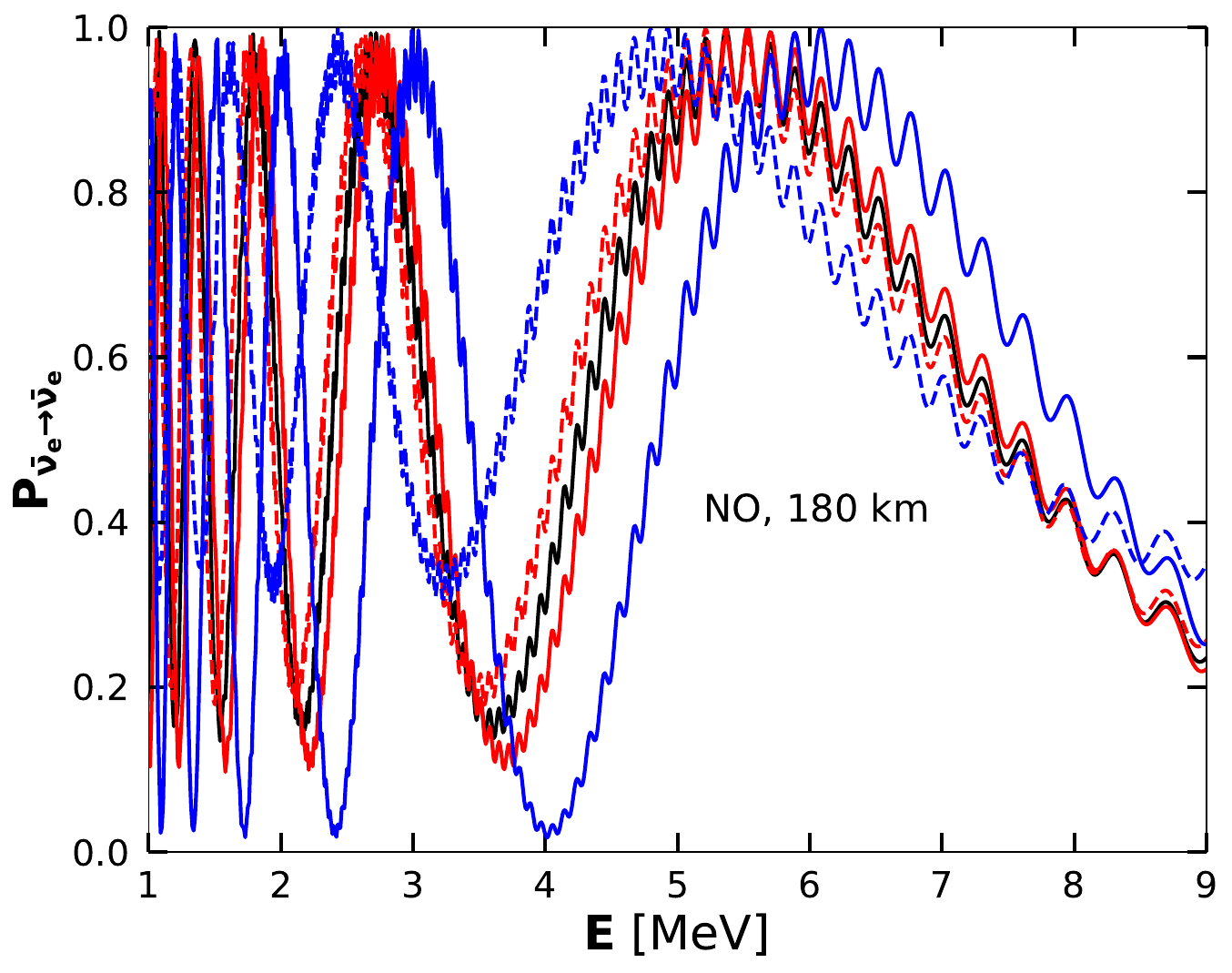}
 \includegraphics[width=.49\textwidth]{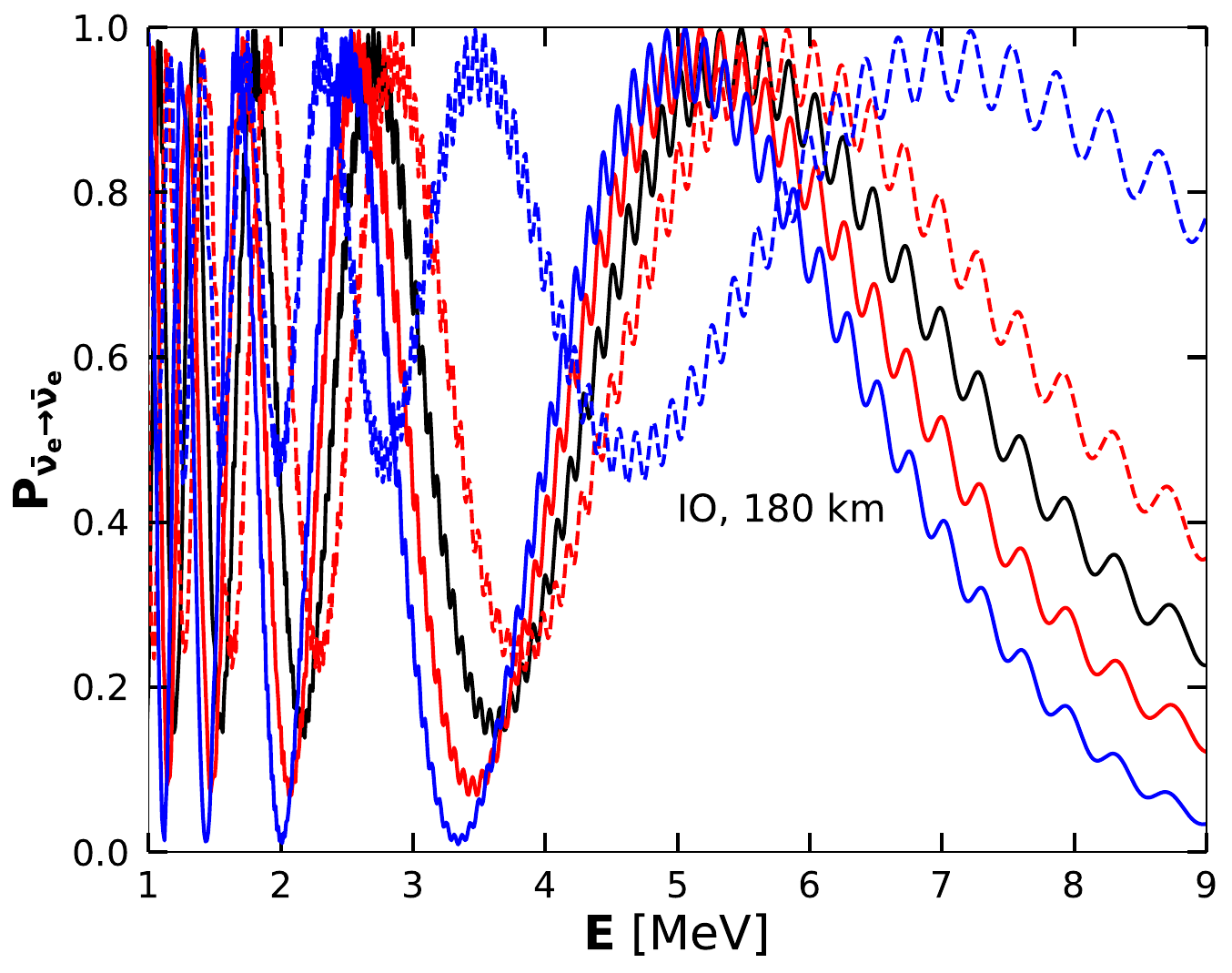}
\caption{\footnotesize The electron antineutrino survival probability $P_{ee}$ including sNSI $\eta_{ee}$ as a function of antineutrino energy for JUNO (top panel) and KamLAND (bottom panel) experiments.  The left(right) panel is for NO(IO). The probabilities for the standard case (marked W/O NSI) are shown in solid black colour. The value of ${\rm M}_{\rm lightest}$ is assumed to be $0.001$ eV. }
\label{fig:oscprob}
\end{figure}

In Fig. \ref{fig:oscprob}, we show the effect of scalar NSI parameter $\eta_{ee}$ on the electron antineutrino disappearance probability $P \left(\bar{\nu}_{e}\rightarrow \bar{\nu}_{e}\right) = P_{ee}$ plotted as a function of the antineutrino energy for JUNO and KamLAND baselines corresponding to $L=53~\rm km$ and $L=180~\rm km$, respectively \footnote{While we present the probability plot for a single baseline of 180 km (KamLAND) to illustrate the impact of the scalar NSI parameter $\eta_{ee}$ on oscillation probabilities for both normal and inverted mass orderings, the event spectra for KamLAND and JUNO account for the smearing effects from all nuclear power plants at varying distances. These effects are also incorporated in the $\chi^2$ calculations for both experiments. In the Appendix \ref{sec:appendix}, we show the effect of sNSI $\eta_{ee}$ at the KamLAND experiment for two more baselines.}. The values of the standard neutrino oscillation parameters have been taken from Table \ref{tab:t1}. The lightest neutrino mass ($m_1$ for NO and $m_3$ for IO) is assumed to be equal to 0.001 eV in all the plots of Fig. \ref{fig:oscprob}. The left panel corresponds to the choice of NO being the true mass ordering  while the right panel corresponds to it being IO. We have used the publicly-available GLoBES package \cite{Huber:2004ka, Huber:2007ji} to generate the oscillation probabilities by appropriately modifying the in built probability functions to incorporate scalar NSI. As mentioned before, our numerical calculations for probabilities match very well with the analytical expressions derived in Sec. \ref{sec:analytical} and also with other works on scalar NSI (sNSI) \cite{Ge:2018uhz, Medhi:2021wxj}. 

For NO, we vary $\eta_{ee}$ in the range $[-0.04,0.04]$ and for IO we vary in the range $[-0.008,0.008]$. As we will see, the magnitudes of these ranges correspond to the values of scalar NSI that give non-negligible deviations in $P_{ee}$ from the standard, three flavour framework expectation (SI). The solid black curve in each plot of Fig. \ref{fig:oscprob} corresponds to SI. Probabilities corresponding to the positive (negative) values of sNSI parameter in each plot are shown with solid (dashed) curves. From \Fig \ref{fig:oscprob}, we make the following observations. 

\begin{itemize} 

\item We observe two discernible oscillatory features in $P_{ee}$ for the given choice of energy ranges and baseline. The longer-wavelength oscillations are $\Delta m^2_{21}$-driven while the smaller-wavelength oscillations are $\Delta m^2_{31}$-driven. As argued in Sec. \ref{sec:analytical}, the effect of sNSI on $\theta_{13}$ is small. If we only look at the change in the $\Delta m^2_{21}$-driven oscillations, we see that in each plot there is an energy at which the probabilities for positive and negative $\eta_{\alpha\alpha}$ coincide. For example, in the case of $\eta_{ee}$ and NO, this occurs at $E_{\bar{\nu}}\approx 2~\rm MeV$ for JUNO while for KamLAND this occurs at $E_{\bar{\nu}}\approx 3~\rm MeV$ and $6~\rm MeV$. The enhancement or suppression of $P_{ee}$, compared to the standard case, as a function of the sign of $\eta_{ee}$ flips at this energy value. 

\item The effect of positive (negative) \ee is to suppress (enhance) $P_{ee}$ compared to the SI case for JUNO. This behaviour of \ee is observed for both NO and IO and can be seen in the top panels of \fig \ref{fig:oscprob}.  



\item An important feature to be noticed is that significant deviations are observed in $P_{ee}$ in the case IO with sNSI values an order of magnitude lower compared to what we observe for NO. Thus, we expect JUNO to impose stronger constraints on diagonal sNSI for IO. 

\end{itemize}


\section{The KamLAND and JUNO experiments} 
\label{sec:expt} 

The Kamioka Liquid Scintillator Antineutrino Detector (KamLAND) is a neutrino experiment that ran from 2002 to 2012 in Japan. Its main aim was to observe the oscillation of electron anti-neutrinos when they travel distances of the order of $\sim 200$ km. A kton-size highly purified liquid scintillator detector observed inverse-beta decay events due to anti-neutrinos arriving from 16 nuclear power plants situated at distances ranging from $140$ km to $215$ km. For the purpose of this study, we have considered the data presented in \cite{KamLAND:2013rgu}. Information on background events and systematic uncertainties are also taken from \cite{KamLAND:2013rgu}. The main backgrounds for KamLAND come from geo-neutrinos. For signal events, we consider $5\%$ systematic error and $2\%$ energy calibration error. For background events, we consider $20\%$ systematic error and $2\%$ energy calibration error.

The Jiangmen Underground Neutrino Observatory (JUNO) experiment \cite{An:2015jdp}  is a multi-national neutrino experiment based in China. The construction phase is nearing completion and it is expected to start taking data in 2025. It aims to observe reactor antineutrinos from several nuclear power plants located at Yangjiang and Taishan. It will consist of a 20 kton fiducial mass liquid scintillator detector situated at an average baseline of approximately $53$ km from the reactors. This detector is projected with the ability to reconstruct the incoming neutrino energy with an unprecedented resolution $\Delta E/E \sim 0.03/\sqrt{E_{\text{vis}}(\text{MeV})}$ \cite{An:2015jdp}, $E_{\rm vis}$ is the visible neutrino energy.

The principal purpose of JUNO is to measure the NMO. In addition, it will also be able to measure $\theta_{12}$, $\Delta m^2_{21}$ and $\Delta m^2_{31}$ quite precisely. The distant reactors at Daya Bay and Huizhou will also have small contributions of neutrino flux at JUNO but in this work, we ignore these reactor cores as their contribution to signal events is very small. Here, we have taken into account the neutrino sources only at the Yangjiang and Taishan nuclear power plants (with their respective thermal powers and baselines) as mentioned in Table 2 of \cite{An:2015jdp}. The details of backgrounds and systematic errors are adapted from Ref. \cite{An:2015jdp, JUNO:2021vlw}. The main backgrounds come from geo-neutrino events at low energies. We have considered $5\%$ systematic errors for signal and $20\%$ systematic errors for backgrounds. We use $2\%$ energy calibration error for both signal and background. For the analysis presented in this work, we consider a combined signal and background events of around $140,000$. The signals at JUNO are the inverse beta-decay (IBD) events, $\bar{\nu}_{e} + p \rightarrow e^{+} + n$. The bulk of the signal will lie in the energy range $\sim\left[1.8, 8\right]$ MeV. The antineutrino fluxes and IBD cross-sections are relatively well-known \cite{An:2015jdp}. 

The simulation of both KamLAND and JUNO experiments
is performed using GLoBES \cite{Huber:2004ka,Huber:2007ji}. We have also used this software package to generate event plots and the sensitivity results described in Secs. \ref{sec:events}, \ref{sec:results1} and \ref{sec:results2}.

\section{Event rates in the Presence of Scalar NSI}
\label{sec:events} 

\begin{figure}[h]

\centering
\includegraphics[width=0.49\textwidth]{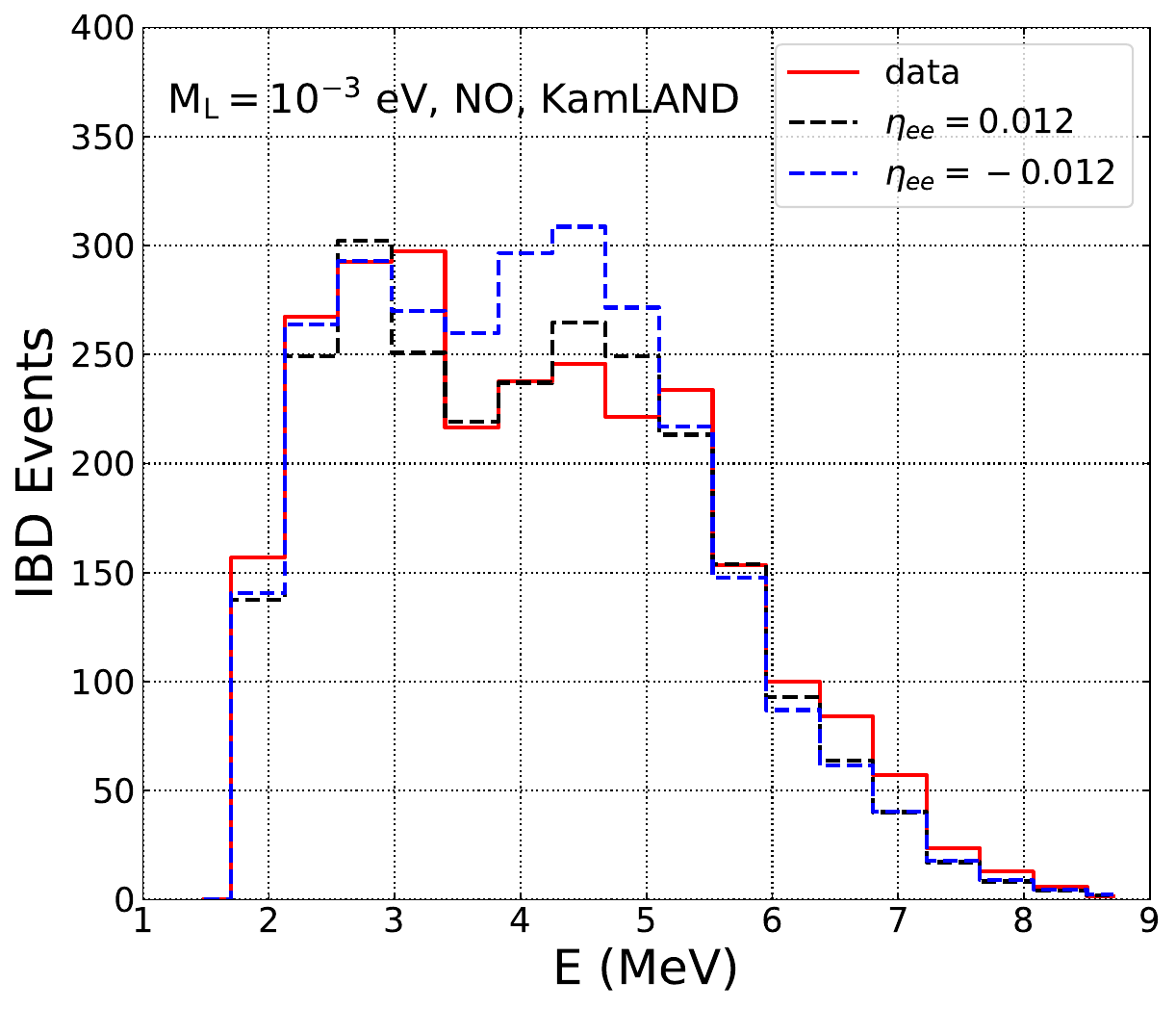}
\includegraphics[width=0.49\textwidth]{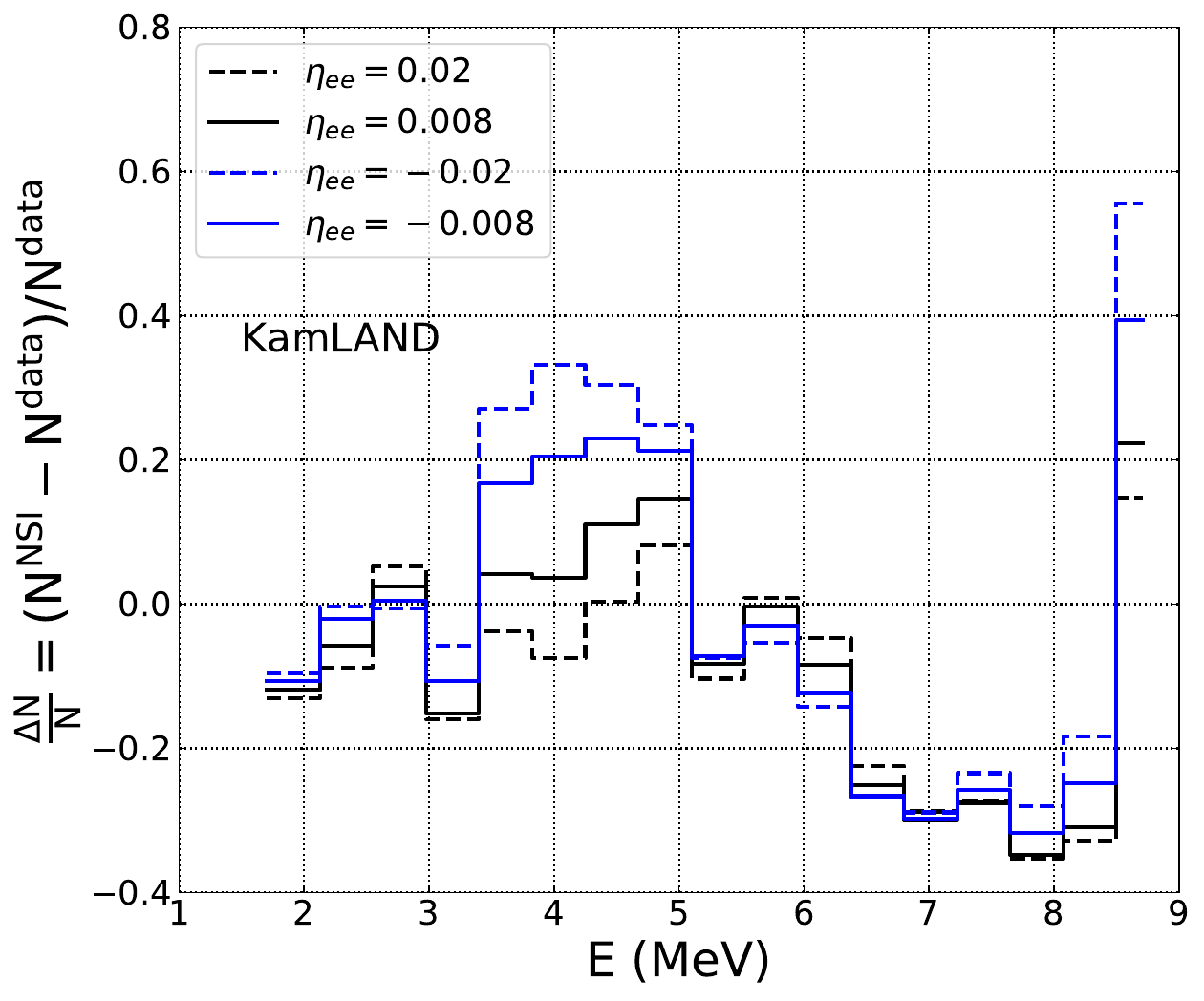}
\includegraphics[width=0.49\textwidth]{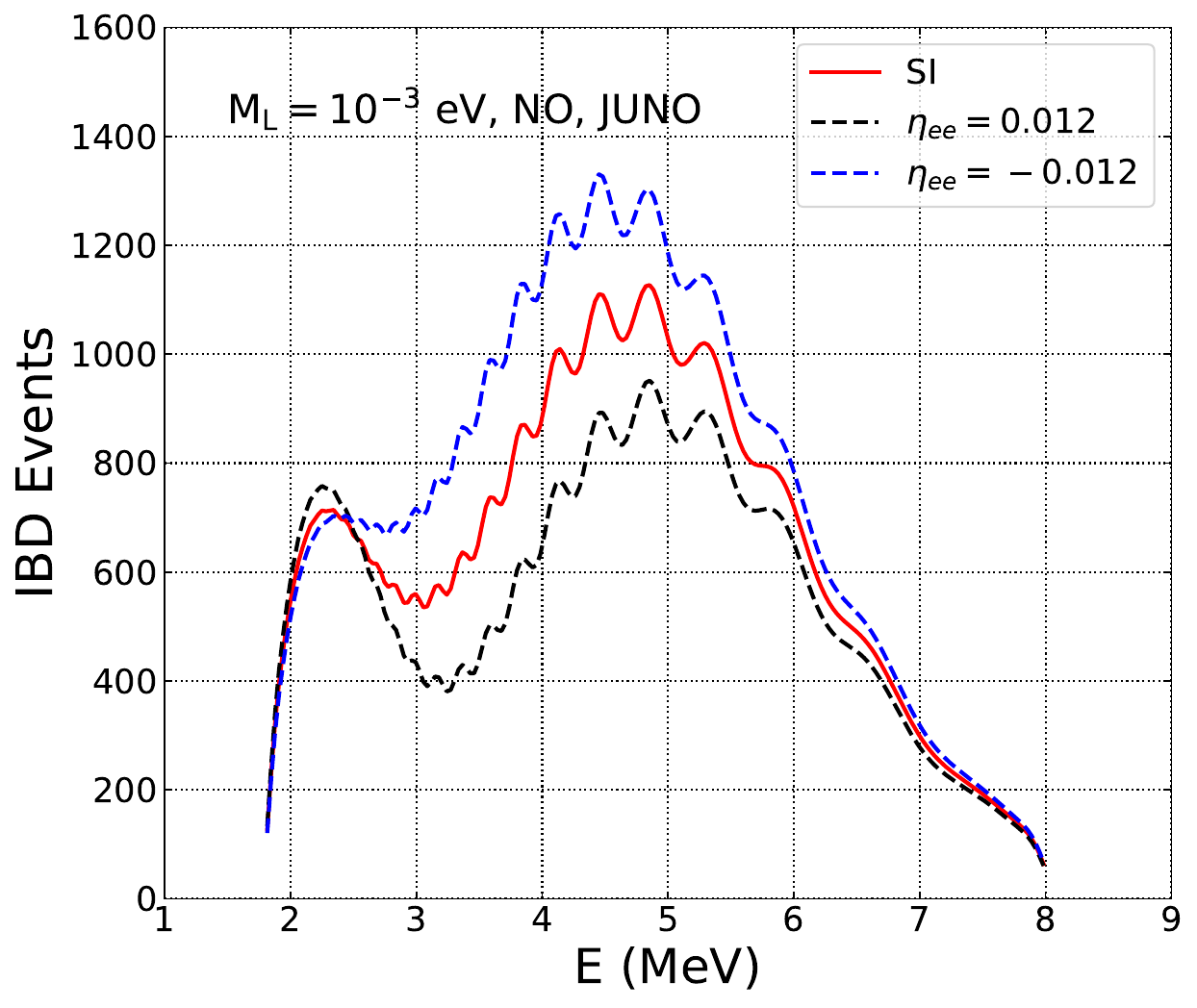}
\includegraphics[width=0.49\textwidth]{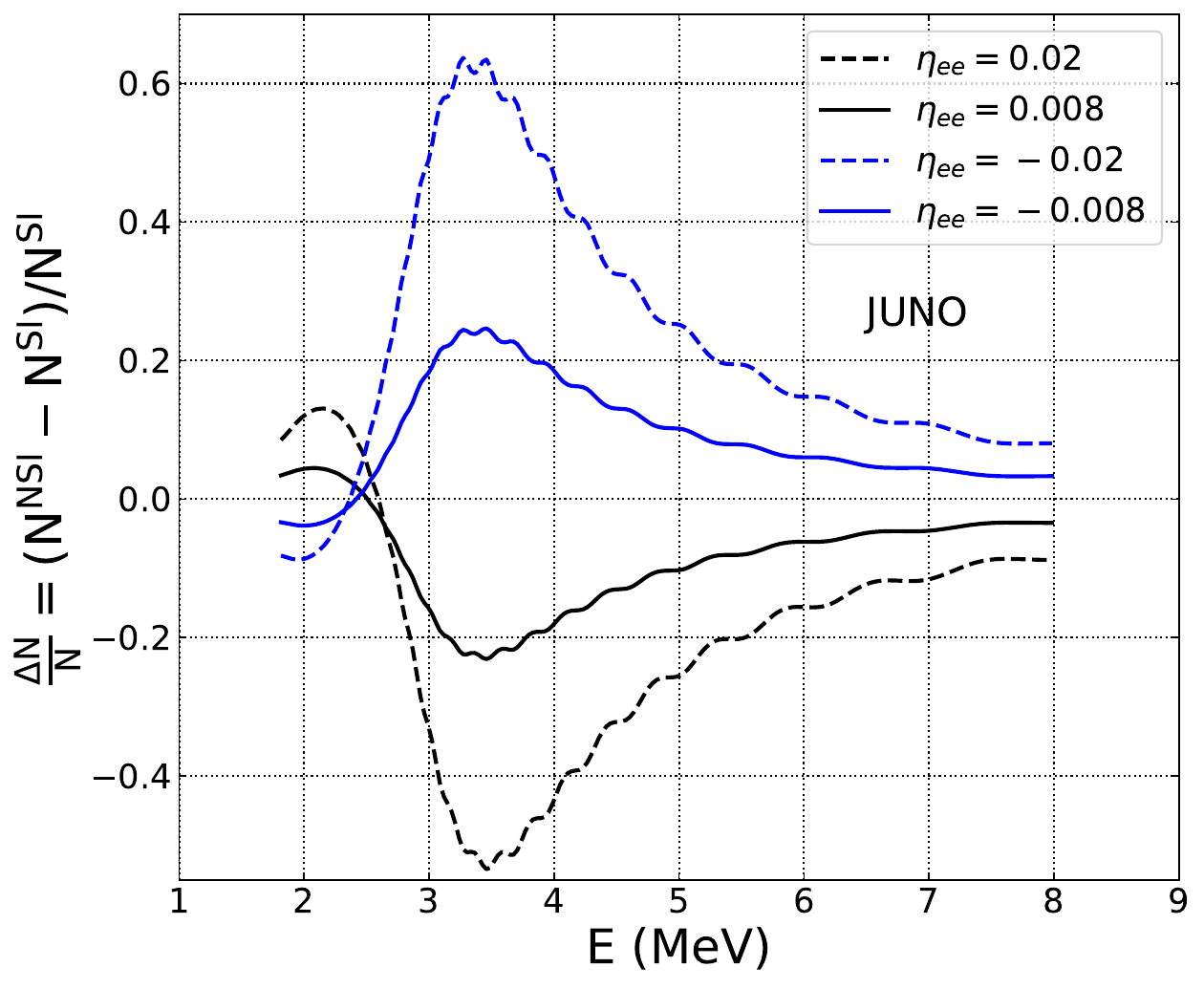}
\caption{\footnotesize Event rates at KamLAND (top panel) and JUNO (bottom panel) assuming NO in the presence of sNSI. The left panel shows the events rates for the standard case, and for $\vert\eta_{ee}\vert = 0.012$. The right panel represents the relative change in event rates with sNSI compared to the standard case for $\vert\eta_{ee}\vert = 0.008, 0.02$.}
\label{fig:eventsNO}
\end{figure}
 
In this section we present the simulated event rates at KamLAND and JUNO in presence of sNSI and check that the features of the event spectra are in congruence with those observed in the probability plots in \fig \ref{fig:oscprob}. For KamLAND, we follow the reconstructed energy binning as done in \cite{KamLAND:2013rgu}; while for JUNO we consider reconstructed antineutrino energy in the range $\left[1.8, 8.0\right]$ MeV divided into 200 bins of $0.031$ MeV bin-width. We show results for different values of $\eta_{ee}$. We assume it to be real and set the remaining diagonal and off-diagonal sNSI to be 0. We choose the mass of the lightest mass eigenstate to be $10^{-3}~\rm eV$. For NO, the lightest mass eigenstate is $m_{1}$ while for IO it is $m_{3}$. The values of standard oscillation parameters are as shown in Table \ref{tab:t1}.

\begin{figure}[h]
\centering
\includegraphics[width=0.65\textwidth]{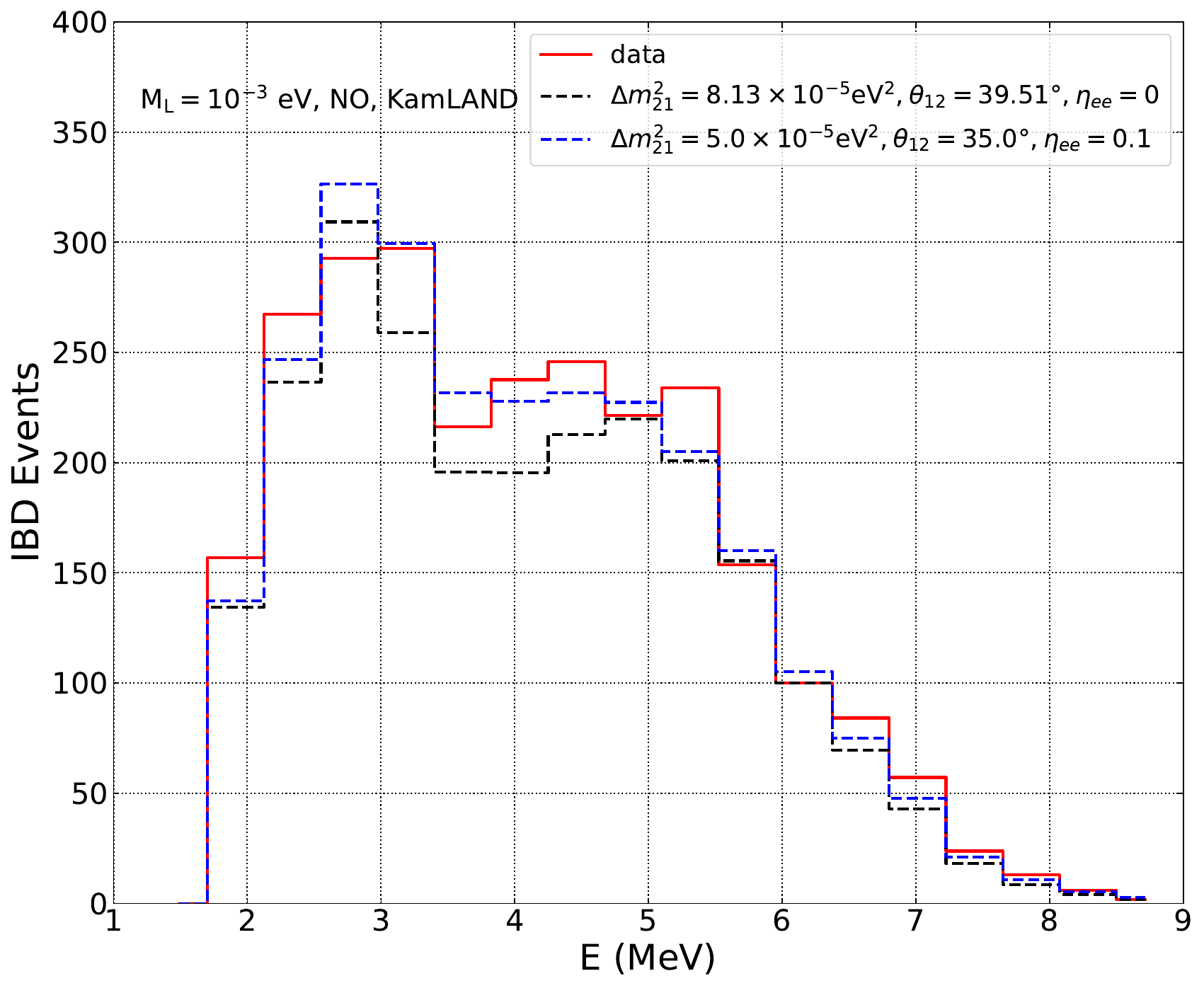}
\caption{\footnotesize Event spectrum for KamLAND data (in red) along with the $3\sigma$-deviated values of the standard oscillation (in black) and oscillation with sNSI $\eta_{ee}$ (in blue) scenarios. 
}
\label{fig:kamland_deg}
\end{figure}

\begin{figure}[h]

\centering
\includegraphics[width=0.49\textwidth]{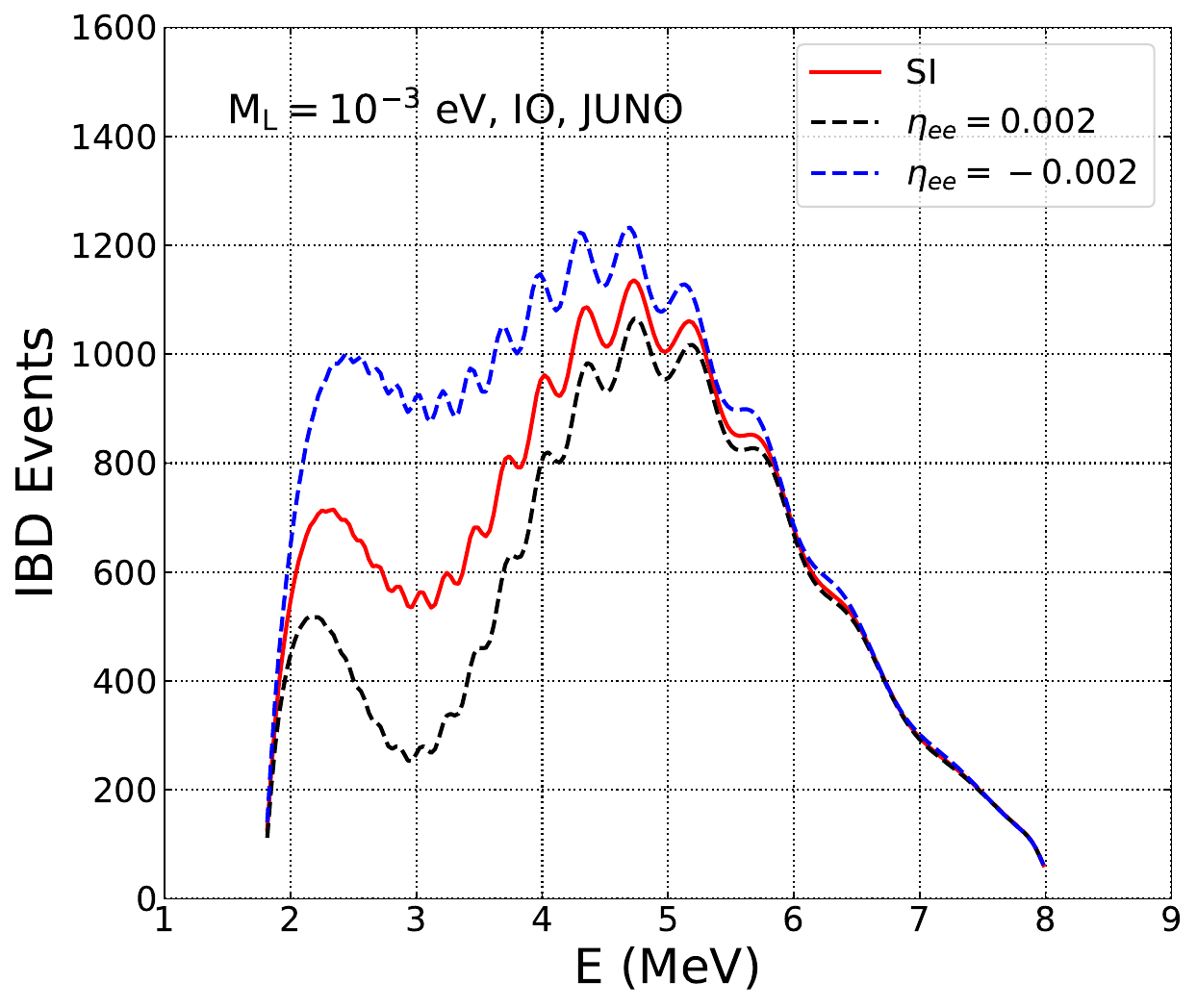}
\includegraphics[width=0.49\textwidth]{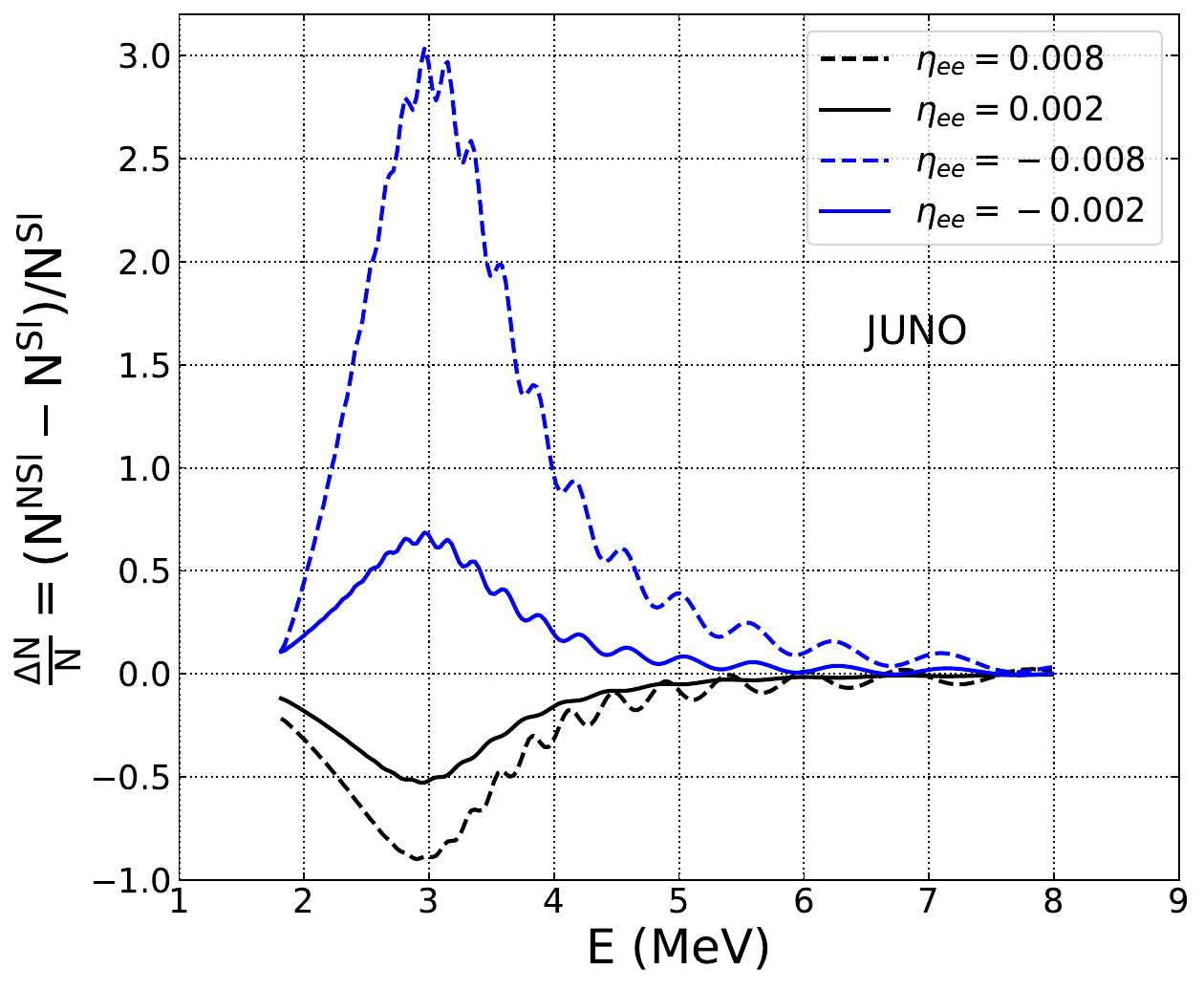}
\caption{\footnotesize Same as in Fig. \ref{fig:eventsNO} but for JUNO and IO case.}
\label{fig:eventsIO}
\end{figure}

The results are shown in Fig. \ref{fig:eventsNO} for NO (for KamLAND and JUNO) and Fig. \ref{fig:eventsIO} for IO (for JUNO only). The left panels in these figures correspond to the total IBD events as a function of neutrino energy for both positive and negative values of sNSI parameter $\eta_{ee}$. The standard event rates/data are shown in red. In the left panel of \fig \ref{fig:eventsNO}, we consider $\vert\eta_{ee}\vert = 0.012$.  
In the right panel of Fig. \ref{fig:eventsNO}, we show the relative difference in the number of expected events per neutrino energy bin for two sets of positive and negative sNSI parameters. These sNSI values correspond to slightly larger and slightly smaller choices compared to the left panel.  Fig. \ref{fig:eventsIO} is generated in a similar way, but for JUNO events assuming IO. 
We can see that change in the value of $\eta_{ee}$ amplifies or suppresses the events rates both in KamLAND as well as in JUNO. Thus, it mimics the change that is caused by varying $\theta_{12}$. We, therefore, anticipate degeneracy in event rates for different combinations of solar oscillation parameters and $\eta_{ee}$. This degeneracy is shown in Fig. \ref{fig:kamland_deg} where KamLAND data (in red) is compared against two set of events spectra - one which corresponds to a choice of $\eta_{ee}=0.0$ (i.e. no sNSI) and $\theta_{12}$ and \sdm $3\sigma$-away (in dashed-black) from the actual best-fit; and another which has $\eta_{ee}=0.1$ along with different values of $\theta_{12}$ and \sdm (dashed-blue). 

This motivates us to 
\begin{itemize}
\item first, fit the KamLAND data assuming the presence of sNSI in theory and calculate the allowed regions of active oscillation parameters. For simplicity, we consider on the $\eta_{ee}$ to be non-zero. 
\item then, estimate JUNO's potential to constrain $\eta_{ee}$ and precisely measure the solar oscillation parameters.
\end{itemize}


\section{Fit to KamLAND data assuming scalar NSI in theory}
\label{sec:results1}

\begin{figure}[h]

\centering
\includegraphics[width=0.65\textwidth]{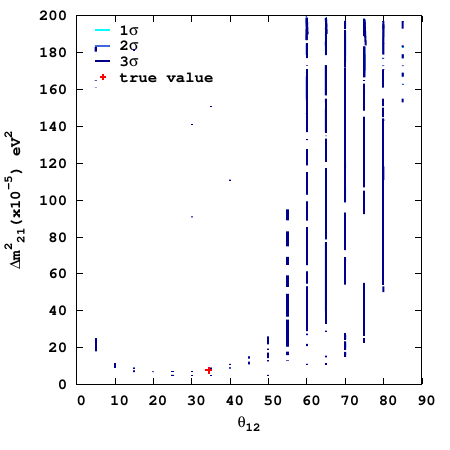}
\caption{\footnotesize Effect of sNSI on measurement of \sdm and $\theta_{12}$ by KamLAND.}
\label{fig:kamland}
\end{figure}

In this section, we fit the KamLAND data assuming the presence of sNSI ($\eta_{ee}$) in the theory. In Fig. \ref{fig:kamland}, we show the $3\sigma$-allowed values of \sdm and $\theta_{12}$. {\it We find that KamLAND data by itself can accommodate $\eta_{ee}$ values ranging from $[-1.0, 1.0]$ at the $3\sigma$ level.} The standard result corresponding to \sdm $= 7.41\times 10^{-5}~\rm eV^2$, $\theta_{12}=33.41^\circ$ and $\eta_{ee} = 0.0$ is shown by the red ``+" point. 
Please note that a direct comparison of our work with earlier literature is difficult, as most other works on this subject assume reactor neutrino experiments to be the benchmark for scalar NSI.  On the other hand, we have assumed sNSI to be true, and our results are model-dependent. Assuming the lightest neutrino mass $(m_{\rm lightest}) ~\sim \mathcal{O} (10^{-5})$ eV, the expected bounds are computed for ESSnuSB~\cite{ESSnuSB:2023lbg} and DUNE~\cite{Singha:2023set} LBL experiments which are $>1$ order of magnitude stronger than the limits obtained from the KamLAND only data. Also note that we have assumed here $(m_{\rm lightest}) ~\sim \mathcal{O} (10^{-3})$ eV. We further emphasize that in this work, we consider only the neutrino data of the KamLAND reactor to derive constraints on sNSI and JUNO for future constraints. The authors in~\cite{Denton:2024upc}, have studied sNSI in the context of solar neutrinos and derived the best constraints on $\eta_{ee}$ (solar) which is around two orders of magnitude stronger than bounds on ($\eta_{ee}$ (Earth)) expected from all the previous works~\cite{PhysRevLett.122.211801, Khan:2019jvr,Smirnov:2019cae,Medhi:2021wxj,Medhi:2022qmu,Sarkar:2022ujy}. This is expected because very large density in the Sun makes solar neutrinos more sensitive to sNSI parameters compared to the terrestrial one. However, in paper~\cite{Denton:2024upc}, the authors have not performed a combined fit to solar + reactor (KamLAND) neutrino data with scalar NSI, rather, they have used priors from KamLAND data. In order to have a clear and complete understanding of sNSI, involving solar neutrino data with KamLAND reactor neutrino data is needed, but we do not attempt that here.


Our results show that the presence of sNSI - the most conservative case where we have considered just one diagonal NSI: $\eta_{ee}$ in the theory - renders the estimation of \sdm and $\theta_{12}$ inconclusive. {\it We find that \sdm as high as $2.0\times10^{-3}~\rm eV^2$ is allowed by the data while $\theta_{12}$ ranging from $5^\circ$ to $85^\circ$ is allowed.} However, if $\theta_{12}$ is close to $33^\circ$ the variation in \sdm is also small in the presence of sNSI. A similar conclusion on the \sdm in the presence of sNSI \ee (solar) is also drawn in the paper~\cite{Denton:2024upc} (see \fig 7) where allowed-region contours are unbounded for $\Delta m^2_{21}$. This is because of the fact that sNSI can exactly play the role of an additional mass term in the Hamiltonian and can give degenerate solutions, making $\Delta \chi^2$ very low. It should also be noted that $\theta_{12}$ is not well-constrained by the reactor neutrino data even in the standard case. Therefore, the addition of one more (sNSI) parameter, namely \ee in the theory, makes $\theta_{12}$ measurement worse by KamLAND-only data. It is shown in~\cite{Denton:2024upc} (see \fig 6) that if \sdm varies between $(10^{-5}-10^{-4})$ eV$^2$, the allowed region for $\theta_{12}$ is small even in the presence of sNSI. But in the present work, we vary both \sdm and $\theta_{12}$ freely and then obtain the allowed regions for both. That is why our allowed regions are significantly large.

\section{Measurements of Solar Oscillation Parameters by JUNO}
\label{sec:results2}

An important physics goal of JUNO is ultra-precise measurement of the solar oscillation parameters \sdm and $\theta_{12}$. It is imperative to ask if the measurement of standard solar oscillation parameters at JUNO is robust against the presence of new physics in the neutrino sector such as sNSI discussed in this work. In this section we estimate how accurately JUNO can reconstruct \sdm and $\theta_{12}$ given that sNSI (considering $\eta_{ee}$ only) exist in nature. 

\begin{figure}[h]

\centering
\includegraphics[width=0.49\textwidth]{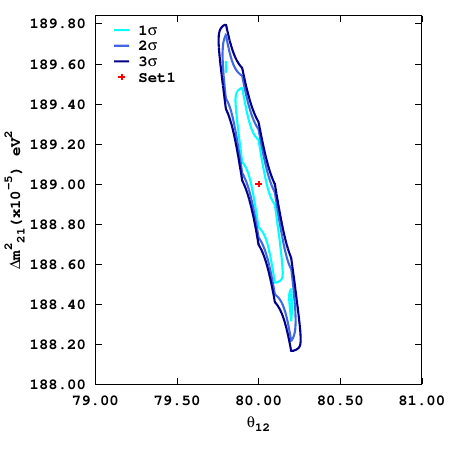}
\includegraphics[width=0.49\textwidth]{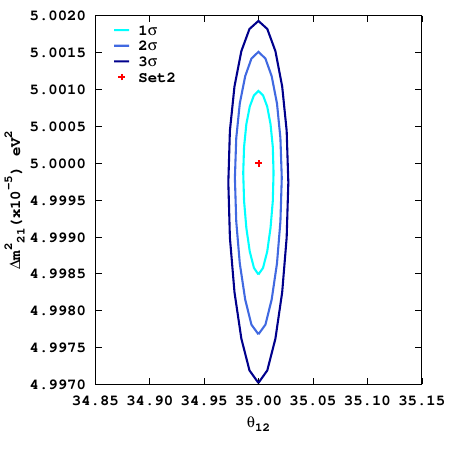}
\includegraphics[width=0.49\textwidth]{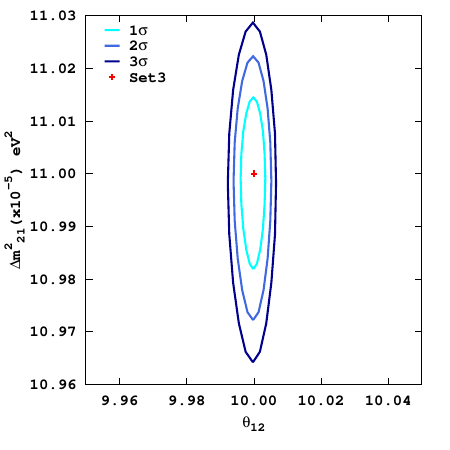}
\includegraphics[width=0.49\textwidth]{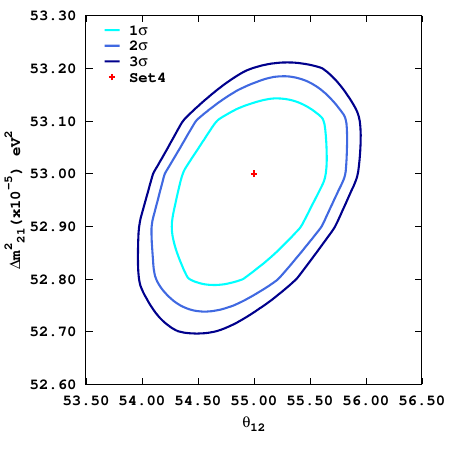}
\caption{\footnotesize Precision measurement of \sdm and $\theta_{12}$ at JUNO in the presence of sNSI.}
\label{fig:th12_dm21}
\end{figure}

To this end, we select some representative combinations of \sdm, $\theta_{12}$ and $\eta_{ee}$ spanning the entire $3\sigma$ allowed ranges from the fit to KamLAND data (as shown in Fig. \ref{fig:kamland}). These values are shown in Table \ref{tab:set_values}. We now consider these four combinations of \sdm, $\theta_{12}$ and $\eta_{ee}$ to be the true case one by one and calculate the allowed regions from JUNO in the test \sdm - $\theta_{12}$ plane. 
The results are shown in Fig. \ref{fig:th12_dm21}. In all the plots the true values are indicated by the red ``+" point. Here we present results for only the normal mass ordering. We have confirmed that marginalizing over test $\theta_{13}$ and test \ldm have minuscule effects on the results; and therefore, we fix these two test parameters at their central values while performing the $\chi^2$-analysis. For each of these plots, we have marginalized over test $\eta_{ee}$ in the range $[-1.0, 1.0]$. 

Our results show that JUNO's ability to constrain \sdm and $\theta_{12}$ remains robust even when we assume the existence of scalar NSI ($\eta_{ee}$) in the theory. Further, the precision could vary from sub-percent to a few percent depending on the true values.  


\begin{table}[h]
    \centering
    \begin{tabular}{|c|c|c|c|}
    \hline
        Input values & \sdm ($\times 10^{-5}$ eV$^2$) & $\theta_{12}$ [deg] & \ee \\
        \hline
         Set1& 189.0 & 80.0 & -1.0\\
         \hline
         Set2& 5.0  & 35.0  & 0.1\\
        \hline
         Set3& 11.0 & 10.0 & 0.25\\
        \hline 
         Set4& 53.0 & 55.0 &-0.45 \\
\hline 
    \end{tabular}
    \caption{True input values of oscillation parameters \sdm, $\theta
_{12}$ and $\eta_{ee}$ randomly chosen from the $3\sigma$ allowed regions from KamLAND in the presence of sNSI.}
    \label{tab:set_values}
\end{table}

\section{Summary and conclusions}
\label{sec:summary}
Precision measurement of neutrino oscillation parameters is one of the main objectives of current and future medium and long-baseline experiments. It becomes important in such a situation to ask if any new physics can affect these measurements. Several works have been done in this regard considering the vector non-standard interactions. In this work, we have studied another simple form of Lorentz structure namely scalar non-standard interactions (sNSI) within the context of reactor neutrino experiments KamLAND and JUNO. We calculate the effective electron antineutrino survival probability considering only one diagonal sNSI element $\eta_{ee}$. Detailed analytical calculations have been performed in this paper to derive an expression for $\bar{\nu}_e\to\bar{\nu}_e$ oscillation probability with scalar NSI in neutrinos where the Hamiltonian matrix due to sNSI is considered diagonal with only one element $\eta_{ee}$.

sNSI appears as a correction to the neutrino mass term in the Hamiltonian. These corrections are independent of neutrino energies and are expressed in terms of the product of $\sqrt{|\Delta m^2_{31}|}$, $\eta_{\alpha \alpha}$ and $\rm m_{lightest}$.
By analyzing $\bar{\nu}_e$ disappearance probability and event spectrum plots for KamLAND and JUNO we find that sNSI impacts the inverted ordering (IO) more profoundly than the normal ordering (NO) of neutrino masses.  We first fit the KamLAND data assuming sNSI ($\eta_{ee}$) in the theory and find that for the choice of lightest neutrino mass $M_L = 0.001$ eV the $3\sigma$ allowed range of $\eta_{ee}$ spans from [-1.0,1.0]. Moreover, sNSI induces significant ambiguity in the determination of solar oscillation parameters \sdm and $\theta_{12}$ by KamLAND only data where values, that are quite different from standard case, are allowed in the presence of sNSI. We choose some of these \sdm and $\theta_{12}$ combinations and test JUNO's ability to measure them assuming the existence of sNSI in nature. 
Our findings suggest that while JUNO will not constrain sNSI parameters any better than KamLAND, its efficacy in constraining \sdm and $\theta_{12}$ remains robust even in the presence of sNSI. Our work emphasizes the crucial need for a comprehensive global analysis of constraints on scalar non-standard interactions before more difficult problems, such as neutrino mass ordering or CP violation in the presence of sNSI are attempted. 

\section*{Acknowledgements}
We thank Andr\'{e} de Gouv\^{e}a and S. Uma Sankar for their useful comments and discussions related to scalar NSI effects in oscillation experiments. We would  also like to thank Alessio Giarnetti for the insightful discussions related to the  Solar neutrino constraints on scalar NSI. 
\appendix 
\section{KamLAND OSCILLATION PROBABILITY FOR OTHER BASELINES}\label{sec:appendix}

\begin{figure}[h]
\centering
\includegraphics[width=.49\textwidth]{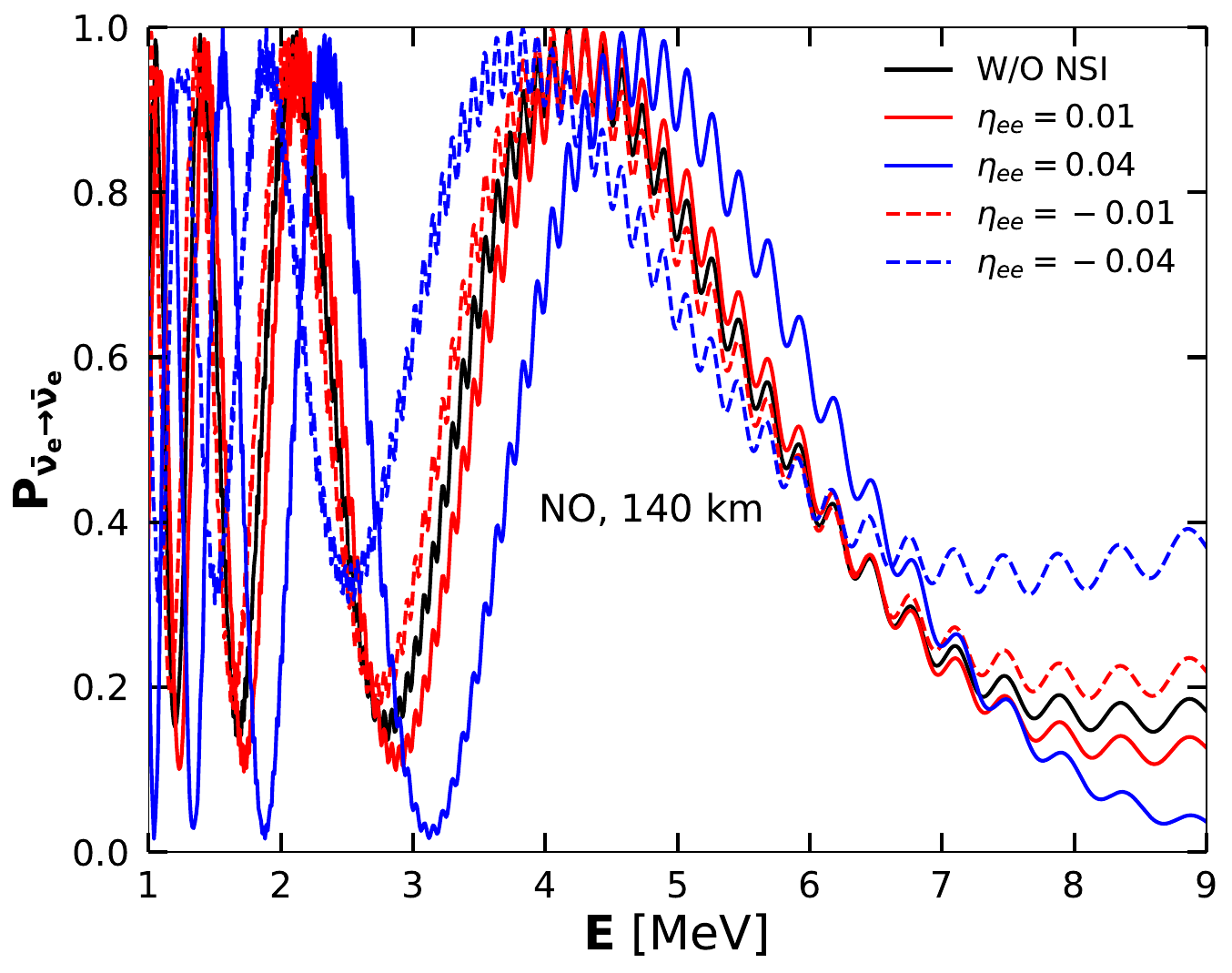}
\includegraphics[width=.49\textwidth]{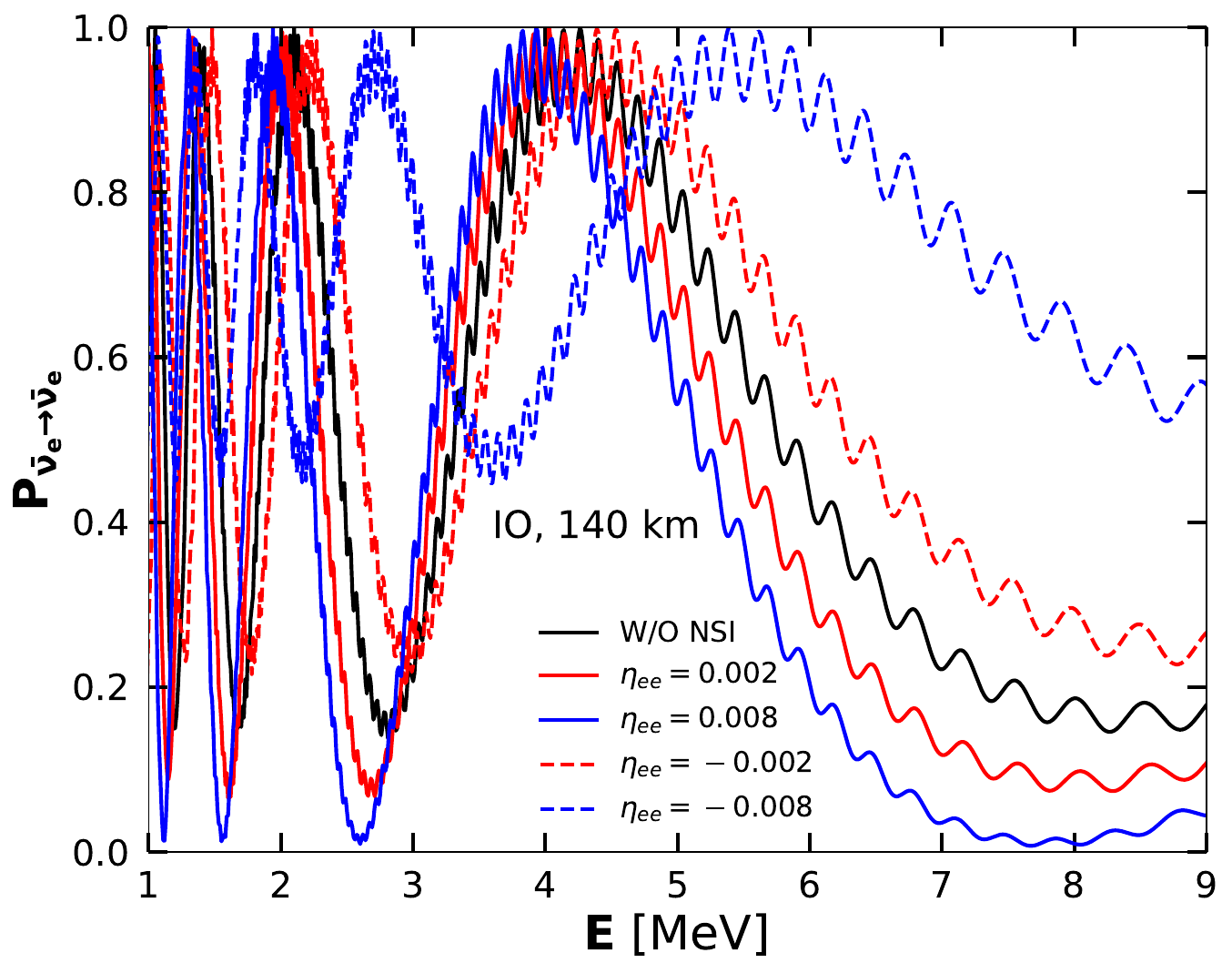}
\includegraphics[width=.49\textwidth]{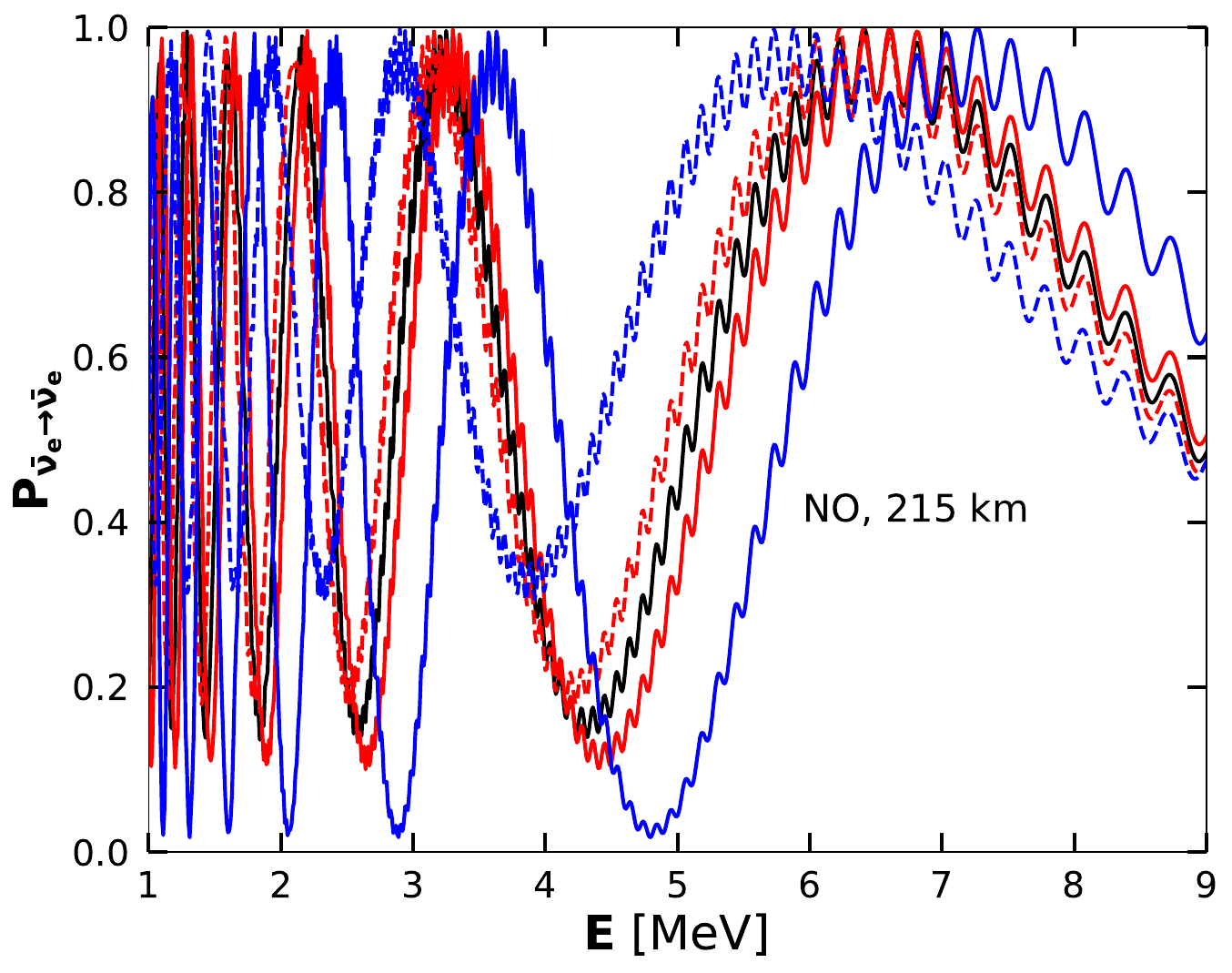}
 \includegraphics[width=.49\textwidth]{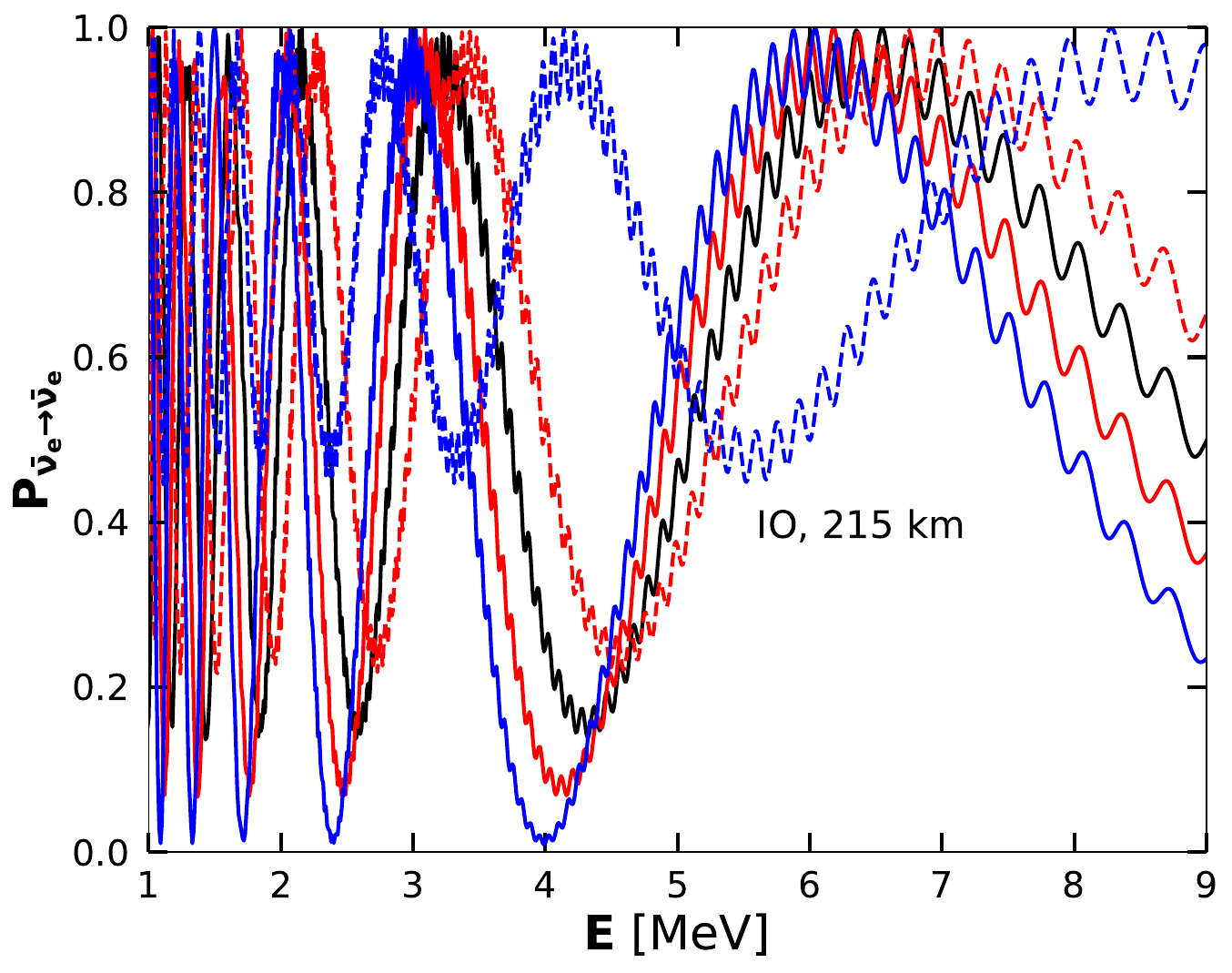}
\caption{\footnotesize The electron antineutrino survival probability $P_{ee}$ including sNSI, $\eta_{ee}$, as a function of antineutrino energy for 140 km (top panel) and 215 km (bottom panel) baselines correspond to the KamLAND experiment.  The left(right) panel is for NO(IO). The probabilities for the standard case (marked W/O NSI) are shown in solid black colour. The value of ${\rm M}_{\rm lightest}$ is assumed to be $0.001$ eV. The discussions presented in Sec. \ref{sec:oscprob} for the KamLAND baseline of 180 km can also be appreciated for other baselines of the KamLAND experiment.}
\label{fig:appendix}
\end{figure}

\bibliography{references.bib,references1.bib}
\end{document}